\documentclass[]{aa}

\usepackage{graphicx,natbib}
\usepackage{floatrow}
\usepackage{subfigure}
\usepackage{amsmath,amssymb}

\usepackage{txfonts}
\usepackage{pifont}

\usepackage{hyperref}

\hypersetup{
   colorlinks=true,
   citecolor=blue,
   linkcolor=blue,
   urlcolor=black
   }

\def\mmw{mean molecular weight }

\def\sma1{semi-major axis}
\def\MR1{mass-radius}
\def\brunt{Brunt-V\"ais\"al\"a }

\def\gl581{Gl\,581\,c}
\def\hd85{HD\,85512\,b}

\def\sizefig{0.56}

\def\ssb{\sigma_\mathrm{SB}}

\def\Rgp{R}

\def\Pr{Pr}
\def\taud{\tilde{\tau}_D}
\def\Rrho{R_\rho}

\def\co2{CO$_2$}
\def\h2o{H$_2$O}
\def\ch4{CH$_4$}
\def\N2{N$_2$}
\def\nh3{NH$_3$}

\def\Fint{F_\mathrm{int}}

\def\grav{g}
\def\cp{c_\mathrm{p}}

\def\hp{H}
\def\temp{T}
\def\vptemp{\theta_\mathrm{v}}
\def\ptemp{\theta}
\def\press{p}
\def\tempm{\bar{T}}
\def\pressm{\bar{p}}
\def\rhom{\bar{\rho}}
\def\delt{\nabla_T}
\def\delad{\nabla_\mathrm{ad}}
\def\delmoist{\nabla_\mathrm{ad}^\star}

\def\delmu{\nabla_\mu}
\def\deltetav{\nabla_{\vptemp}}
\def\delq{\nabla_q}

\def\delrad{\nabla_\mathrm{rad}}

\def\amu{\alpha_{\mu}}

\def \kapt{\kappa_T}
\def\Nt{N_T}
\def\Nmu{N_{\mu}}
\def\Nq{N_{q}}
\def\tNq{\tilde{N}_{q}}

\def\mum{\bar{\mu}}

\def\kapR{\kappa_\mathrm{R}}

\def\freq{\sigma}
\def\wn{k}
\def\mass{m}
\def\massg{m_\mathrm{g}}

\def\pair{p_\mathrm{a}}
\def\rhoair{\rho_\mathrm{a}}
\def\qair{q_\mathrm{a}}

\def\cpair{c_\mathrm{p,a}}
\def\Mair{M_\mathrm{a}}

\def\pvap{p_\mathrm{v}}
\def\rhovap{\rho_\mathrm{v}}
\def\qvap{q_\mathrm{v}}
\def\qvapm{\bar{q}_\mathrm{v}}
\def\dqvap{\d q_\mathrm{v}}
\def\delqvap{\delta q_\mathrm{v}}
\def\qint{q_\mathrm{int}}
\def\qcri{q_\mathrm{cri}}

\def\cpvap{c_\mathrm{p,v}}
\def\Mvap{M_\mathrm{v}}

\def\latent{L}
\def\betacond{\beta_\mathrm{c}}
\def\qsat{q_{\mathrm{s}}}
\def\qsatm{\bar{q}_{\mathrm{s}}}
\def\delqsat{\delta q_{\mathrm{s}}}
\def\gammasat{\gamma_\mathrm{s}}
\def\gammavap{\gamma_\mathrm{v}}
\def\psat{p_{\mathrm{s}}}

\def\qcon{q_\mathrm{c}}

\def\xHetop{x^\mathrm{top}_\mathrm{He}}

\def\dlp{\d \ln \press}

\def\mratio{\varpi}

\def\tcond{\tau_\mathrm{c}}
\def\itcond{\tau_\mathrm{c}^{-1}}
\def\ttcond{\tilde{\tau}_\mathrm{c}}
\def\titcond{\tilde{\tau}_\mathrm{c}^{-1}}
\def\freqD{\chi_D}
\def\freqnu{\chi_\nu}
\def\freqK{\chi_\kappa}
\def\freqDt{\tilde{\chi}_D}
\def\freqnut{\tilde{\chi}_\nu}
\def\freqKt{\tilde{\chi}_\kappa}

\def\d{\mathrm{d}}
\def\i{\mathrm{i}}

\def\dvv{\delta\mathbf{u}}
\def\dv{\delta \upsilon}
\def\vk{\mathbf{k}}
\def\tk{\tilde{k}}
\def\tfreq{\tilde{\freq}}
\def\vr{\mathbf{r}}

\def\uz{\mathbf{\hat{z}}}
\def\vg{\mathbf{g}}
\def\vv{\mathbf{u}}
\def\drho{\delta \rho}
\def\deltemp{\delta \temp}
\def\dmu{\delta \mu}

\newcommand{\balign}[1]{
\begin{align}
#1
\end{align}}

\newcommand{\eq}[1]{Eq.\,(\ref{#1})}
\newcommand{\eqs}[2]{Eqs.\,(\ref{#1}) and (\ref{#2})}
\newcommand{\fig}[1]{Fig.\,\ref{#1}}
\newcommand{\figs}[2]{Figs.\,\ref{#1} and \ref{#2}}
\newcommand{\sect}[1]{Sect.\,\ref{#1}}

\newcommand{\app}[1]{Appendix\,\ref{#1}}
\newcommand{\tab}[1]{Table\,\ref{#1}}

\newcommand{\pd}[2]{\frac{\partial \!\! \ #1}{\partial \!\! \ #2}}

\newcommand{\pdc}[3]{\left. \frac{\partial \!\! \ #1}{\partial \!\! \ #2}\right|_{#3}}

\newcommand{\dd}[2]{\frac{\mathrm{d} \!\! \ #1}{\mathrm{d}\!\! \ #2}}

\newcommand{\ppd}[1]{\partial_{#1}}
\newcommand{\bigd}[1]{D_{#1}\,}

\newcommand{\cmark}{\ding{51}}%
\newcommand{\xmark}{\ding{55}}%

\titlerunning{Inhibition of convection by condensation}
\authorrunning{Leconte et al.}

\begin{document}

\title{
Condensation-inhibited convection in hydrogen-rich atmospheres
}
\subtitle{Stability against double-diffusive processes \\and thermal profiles for Jupiter, Saturn, Uranus, and Neptune
}

\author{J\'er\'emy Leconte\inst{1}, Franck Selsis\inst{1}, Franck Hersant\inst{1}, and Tristan Guillot\inst{2}}

\institute{
Laboratoire d'astrophysique de Bordeaux, Univ. Bordeaux, CNRS, B18N, all\'ee Geoffroy 
Saint-Hilaire, 33615 Pessac, France
\and
Laboratoire Lagrange, UMR 7293, Universit\'e de Nice-Sophia Antipolis, CNRS, Observatoire de la C\^ote d'Azur, 06304 Nice Cedex 4, France
}

\date{}

\offprints{jeremy.leconte@u-bordeaux.fr}

\abstract{
In an atmosphere, a cloud condensation region is characterized by a strong vertical gradient in the abundance of the related condensing species. On Earth, the ensuing gradient of mean molecular weight has relatively few dynamical consequences because N$_2$ is heavier than water vapor, so that only the release of latent heat significantly impacts convection. On the contrary, in an hydrogen dominated atmosphere (e.g. giant planets), all condensing species are significantly heavier than the background gas. This can stabilize the atmosphere against convection near a cloud deck if the enrichment in the given species exceeds a critical threshold. This raises two questions. What is transporting energy in such a stabilized layer, and how affected can the thermal profile of giant planets be? To answer these questions, we first carry out a linear analysis of the convective and double-diffusive instabilities in a condensable medium showing that an efficient condensation can suppress double-diffusive convection. This suggests that a stable radiative layer can form near a cloud condensation level, leading to an increase in the temperature of the deep adiabat. Then, we investigate the impact of the condensation of the most abundant species---water---with a steady-state atmosphere model. Compared to standard models, the temperature increase can reach several hundred degrees at the quenching depth of key chemical tracers. Overall, this effect could have many implications for our understanding of the dynamical and chemical state of the atmosphere of giant planets, for their future observations (with \textit{Juno} for example), and for their internal evolution.
}


\maketitle

\section{Clouds and convection in giant planets}

Evidence for a significant volatile enrichment of the interior of giant planets has been accumulating for some time now \citep[see][and reference therein]{Ste82rev,Gui99}. And recent data seem to point at elemental abundances in C, N, and O that might be even more supersolar than previously thought \citep{NAC96}. But because of their old age and distance from the Sun, giant planets' upper atmospheres are cold enough to condense the main usual carriers of these species in a hydrogen rich environment (e.g. \h2o, \nh3, and \ch4) which are thus present in trace amounts only.

This tells us that all giant planets' atmospheres harbor regions where the chemical composition of the gas changes and where clouds form---the very clouds which give giant planets their rich colors. But what is the effect of such a composition gradient on the thermal properties of the atmosphere? To this question, our first answer, based on our experience with the Earth's atmosphere, is often that condensation of volatiles releases latent heat which facilitates convection. As a result, in an unstable, saturated region, the temperature profile follows a subadiabatic thermal gradient, the so-called "moist adiabat."

Meanwhile, it is often forgotten that this composition gradient also entails a gradient in the mean molecular weight of the gas which can affect the thermal profile \citep{Sto86,Gui95}. This oversight is mostly due to the fact that in our own atmosphere, the condensable species, water, is lighter than the background atmosphere, and not by a large factor. Thus, an already convectively unstable medium is only slightly further destabilized.

In hydrogen dominated atmospheres, however, almost any condensable species would be significantly heavier than the uncondensable background gas. As has been shown by \citet{Gui95} in the limit without diffusion, and demonstrated in the general case hereafter, this can lead to the stabilization of the atmosphere against convection. The reason is that if an eddy rises following the moist adiabat in a superadiabatic region, its temperature will be slightly larger than its surrounding medium. Usually this would entail a lower density and a positive buoyancy which would cause the eddy to keep rising, i.e. that the medium is convectively unstable. But in our case, the higher temperature of the eddy also means that it is able to retain more vapor\footnote{Hereafter, any condensable species in its gaseous form will be referred to as "vapor."}, so that the mean molecular weight in the eddy is larger than in the surrounding gas. In other words, the abundance of the condensable species drops more slowly in the rising parcel than in the environment. This can potentially cause the density of the eddy to be higher and thus its buoyancy to be negative.

Under such conditions, overturning, large-scale convection would be inhibited. As a result, the efficiency of energy transport should be greatly reduced, and the thermal gradient in such a layer could be significantly superadiabatic: the temperatures below this layer could be significantly higher than expected. This raises two important questions.

\subsection{What is transporting energy in such a condensation-stabilized layer?}

The importance of the temperature increase depends on exactly how thick and how superadiabatic such a stable layer could be. This, in turn, depends on the process that will transport energy in these stable regions, the identification of which will be one of the two main goals of our study. 

If convective motions are completely shut down, one might expect that radiative processes would take over. Then, the thermal gradient in the stable layer would equal the gradient needed for radiation to carry away all of the outgoing internal flux, the so-called radiative gradient ($\delrad$).

But we know that when a fluid exhibits a gradient of mean molecular weight and that heat is allowed to diffuse, an otherwise stable fluid can be unstable to the double-diffusive instability \citep{Ste60}. Then turbulent motion develops in the fluid which enhances the energy transport and, consequently, reduces the thermal gradient compared to the fully stable case \citep{Ste79,RGT11}. 

However, most studies of the double-diffusive instability restrain themselves to the case of a mixture of non-condensable fluids \citep{RGT11,MGS12}. While this is completely valid when treating the transport of salt in the oceans, and might be in some astrophysical contexts, it surely is not in the present context, where the gradient of \mmw itself is produced by condensation.

In a first attempt at solving this question, in \sect{sec:linear_anlysis}, we develop a linear analysis of the double-diffusive instability including the effect of condensation. To that purpose, we complement the usual set of Navier Stockes equations in the Boussinesq approximation with an equation of state for a two-phase fluid, one of which condenses. 
We verify that this set of equation adequately captures the usual double-diffusive instability for a non-condensable species in the limit where condensation is inefficient.

Then, in \sect{sec:inhibition}, we demonstrate that, in the expected regime where condensation is almost instantaneous compared to other processes, the double-diffusive instability is killed by the condensation. A simple reason for this is that the distribution of the vapor becomes controlled by the saturation vapor pressure which is itself controlled by the temperature. Thus both heat and solute effectively diffuse with the same diffusivity whereas heat needs to diffuse more rapidly to trigger the usual double-diffusive instability. 
As a result, if there is no other source of turbulence inside the condensation layer, we show that energy should be only transported by radiation, and the thermal gradient should be close to the radiative one.

\subsection{Potential implications for giant planets?}

Because water condensation---water being the species with the most important potential effect as will be discussed below---occurs below the region where temperature data are available for the four Solar System major planets, such an effect would have been disregarded in previous modeling of their deep atmosphere. This could have various dramatic implications.
\begin{itemize}
\item[$\bullet$] Accounting for a previously ignored, significantly superadiabatic layer would result in raising our estimates of both the present heat content and volatile enrichment of the planet \citep{Gui05,LC12}.
\item[$\bullet$] The reduced efficiency of the heat transport in the atmosphere would also modify the cooling history of the interior \citep{GGC94,GCG95,CB07,LC13}.
\item[$\bullet$] A higher temperature at depth would affect the thermochemical balance in the atmosphere, thus changing the various compositional profiles and the link between the abundance of trace gases in the atmosphere and the deep elemental abundances \citep{MBL00,CML14}.
\item[$\bullet$] Our ability to retrieve of the deep abundance of some species strongly depends on our knowledge of the temperature profile in these regions \citep{BS89,CPG15}. This is especially important for future space missions like \textit{Juno} \citep[][]{PJO08,DSD14}.
\item[$\bullet$] A highly stable layer would modify the vertical dynamics of the atmosphere, possibly causing some of the observed giant storms \citep{LI15}. It would also create a wave duct for the propagation of gravity waves \citep[e.g.,][]{Ingersoll+Kanamori1995}.
\item[$\bullet$] Generally, the presence and extent of stable regions have an important role in the establishment of the tropospheric jets and their propagation to the deeper atmosphere \citep{SGL06}.
\end{itemize}
If this applies to Solar System giants, it of course also applies to any exoplanet with a hydrogen-rich atmosphere which is substantially enriched in condensable species, and cool enough for these species to condense.

To be able to quantify some of these effects in more detail, we numerically integrate atmospheric  profiles for the four Solar System giant planets for various plausible volatile enrichments (especially water). We recover the fact that the stabilizing effect of the \mmw in Saturn can indeed create a stable layer, as first shown by \citet{LI15}, but further demonstrate that this can indeed change the temperature of the deep adiabat. Furthermore, in Uranus and Neptune, the superadiabatic layer created by water condensation can cause an increase of several hundreds of kelvins below a few hundred bars. 

\section{Convection in presence of a condensible species in the adiabatic limit: basic concepts}\label{sec:conv_adiabatic}

\subsection{Dry processes}\label{sec:dry_processes}

Even when a condensible species is present, convection can still sometimes occur without any condensation. This for example happens when the gas is not locally saturated in vapor and will be hereafter called \textit{dry convection}. 

For a gas at temperature $T$ and pressure $p$, it is well known that, when thermal diffusivity and viscosity can be neglected, the criterion for dry convection to arise is that the density of a lifted parcel of air decreases faster than the density of the surrounding air. This usually gives rise to the \citet{SH58} criterion, \balign{\delt>\delad,}
where $\delt\equiv\dd{\ln T}{\ln p}$ is the thermal gradient, and $\delad\equiv\pdc{\ln T}{\ln p}{\mathrm{ad}}$ the adiabatic gradient. But in the presence of a variable species whose distribution is inhomogeneous, the \mmw of the gas, $\mu$, can vary, and the destabilizing super adiabaticity must now be large enough to counteract the stabilizing effect of a \mmw gradient. In mathematical terms, with $\delmu\equiv\dd{\ln \mu}{\ln p}$, the  \citet{Led47} criterion reads
\balign{\label{ledoux}
\delt>\delad+\delmu}
for a perfect gas\footnote{The general formula is $\delt>\delad-\frac{\pdc{\ln \rho}{\ln \mu}{p,T}}{\pdc{\ln \rho}{\ln T}{p,\mu}}\delmu$}. It is very important to note that $\delmu$ appears only on one side because the composition is kept fixed in the moving parcel, i.e. we have a \textit{dry} process where no condensation/vaporization occurs. The situation will be very different in the next section. 

A result that is a little less known to the astrophysical community is that, for a perfect gas, this criterion allows us to define a very handy quantity---the so-called \textit{virtual potential temperature} \citep{Cur03}.
Consider a medium made of a non-condensable phase (or air, denoted with a subscript a), a condensable gas (or vapor, denoted by v) and condensed material (denoted by c) assumed to rainout instantaneously. The equation of state writes
\balign{
\press\equiv\pair+\pvap \equiv \rhoair \frac{\Rgp}{\Mair} \temp +\rhovap \frac{\Rgp}{\Mvap} \temp\equiv \rho \frac{\Rgp}{\mu}\temp, }
where $p_\i$, $\rho_\i$, and $M_\i$ are respectively the pressures, densities, and molar masses of each gas; $\Rgp$ being the molar ideal gas constant. The density of each gaseous species is defined as the ratio of the mass of the species over the total volume of the gas parcel ($\rho_\i\equiv \mass_\i/V$). Finally, we define the mass mixing ratio of any component as the ratio of its mass over the mass of gas ($q_\i\equiv\mass_\i/\massg\equiv \rho_\i/(\rhoair+\rhovap)$). This convention entails $\qair+\qvap=1$. The mean specific heat capacity is given by $\cp=\cpair+\qvap (\cpvap-\cpair)$ and the mean molecular weight by
\balign{\label{meanmolweight}
\frac{1}{\mu}=\frac{1-\qvap}{\Mair}+\frac{\qvap}{\Mvap}\Rightarrow \frac {\d \ln \mu}{\d \ln \qvap}=\mu\, \qvap\left(\frac{1}{\Mair}-\frac{1}{\Mvap}\right)\equiv \amu 
.}
Introducing the reduced mean molar mass difference, $\mratio~\equiv~(\Mvap-\Mair)/\Mvap$, we get
\balign{\label{defamu}\amu= \frac{\mratio \qvap}{1-\mratio \qvap},\  \mathrm{and,}\ \frac{1}{\mu}= \left(1-\mratio \qvap \right) \frac{1}{\Mair}}
where $\amu$ quantifies the change in relative buoyancy due to a change in vapor mixing ratio.  

Using these notations, one can see that the Ledoux criterion simply writes\balign{\delt-\delad-\delmu=\dd{ \ln \vptemp }{\ln \press}\equiv{\deltetav}>0 \label{stabciteriondry}}
where
\balign{
\vptemp&\equiv\temp  \left(1- \mratio \qvap\right) e^{-\int_{\press_0}^\press \frac{\Rgp}{\mu \,\cp}\, \d \ln \press} ,}
is the \textit{virtual potential temperature}. When the average heat capacity and \mmw are constant in the medium it can be analytically integrated, yielding
\balign{
\vptemp=\temp \left(1- \mratio \qvap\right)\left(\frac{\press_0}{\press}\right)^{\frac{\Rgp}{\mu \,\cp}} .}
The advantage of this quantity is that it integrates two effects.

First, it integrates the compressibility of the gas through the pressure factor so that a homogeneous medium is convective if its \textit{potential temperature}, \balign{\label{pottemperature}\ptemp&\equiv\temp e^{-\int_{\press_0}^\press \frac{\Rgp}{\mu \,\cp}\, \d \ln \press}\\
&\approx \temp\left(\press_0/\press\right)^{\frac{\Rgp}{\mu \,\cp}},} increases with depth. The potential temperature is the temperature that the gas would have if adiabatically displaced to an arbitrary level\footnote{In the following, $\press_0$ will always be taken equal to 1\,bar.} with pressure $\press_0$.

Second, it accounts for the change in \mmw due to the variable amount of vapor. Indeed, the virtual potential temperature is the potential temperature that a parcel of dry air should have to have the same density (at a given pressure) as the one of the actual moist air parcel, $\vptemp=\ptemp\,\Mair/\mu.$
 One can directly see that for almost any condensible species in a hydrogen dominated atmosphere, $\Mvap>\Mair$, so that increasing the vapor content amounts to decreasing the virtual temperature (increasing the density) of the gas, contrary to what happens on Earth. 

\subsection{Moist processes}\label{sec:moist_processes}

\subsubsection{Latent heat effect and the \textit{moist adiabat}}

When saturation is reached, a rising parcel will undergo condensation of part of its vapor phase. This releases latent energy that tends to heat the gas and contributes to the positive buoyancy of the parcel. In other words, this facilitates convection. To formalize this, the thermodynamic properties of the vapor are described by its specific vaporization/sublimation latent heat, $\latent$, and its saturation pressure curve, $\psat(T)$, or equivalently, $\qsat(T,\press)$, the saturation mass mixing ratio. Thus, the gradient of vapor mixing ratio at saturation at constant pressure,
\balign{\label{defgammasat}
\gammasat\equiv\pdc{\ln \qsat}{\ln T}{\press}= \left(1-\mratio \qsat \right)\dd{\ln \psat}{\ln T}= \left(1-\mratio \qsat \right)\frac{\Mvap\latent}{\Rgp \temp},
}
is also known as a function of temperature and pressure (See \app{app:relations} for details).

With these notations, it is a well known result that, when \mmw effects are \textit{disregarded}, the atmosphere undergoes moist convection if it is saturated and if the thermal gradient exceeds the \textit{moist adiabat} given by
\balign{\label{moistadiabat}
\delmoist\equiv    \frac{\Rgp}{\mu\,\cp} \left(1+\frac{\qsat}{1-\qsat} \frac{\Mair \latent}{\Rgp \temp}\right)/\left(1+\frac{\qsat}{1-\qsat} \frac{\latent}{\cp \temp}\gammasat\right),
}
where we recognize the usual dry adiabat for a perfect gas, $\delad~=~\Rgp / (\mu\, \cp)$. This formula accounts for the fact that water may not be a trace gas \citep{Pie10,LFC13b}.

\subsubsection{Mean molecular weight effect}\label{sec:conv_adiabatic_mmw}

When the \mmw is taken into account, things get a little more complex. Indeed, as the parcel rises, condensation will occur and the \mmw of the gas will change. But for moist convection to occur, the environment must be saturated so that the vapor mixing ratio also changes with altitude outside the parcel. In the end, one must compute the differential effect in both $\temp$ and $\mu$, knowing that the two are related by the Clausius-Clapeyron relation.

Let us start by computing the gradient of $\mu$ for a saturated medium with an arbitrary thermal gradient $\delt$. This yields
\balign{\label{delmudelt}
\delmu=\dd{\ln \mu}{\ln \qvap} \dd{\ln\qvap}{\ln \press}&=\dd{\ln \mu}{\ln \qvap} \left[\pdc{\ln \qsat}{\ln \temp}{\press}\dd{\ln \temp}{\ln \press}+\pdc{\ln \qsat}{\ln \press}{\temp}\right]\nonumber\\
&=\amu\left[\gammasat \delt -\left(1-\mratio \qsat\right)\right]\nonumber\\
&=\mratio \qsat \left[\dd{\ln \psat}{\ln \temp} \,\delt -1\right], }
where the second equality assumes that the medium stays saturated, and the others used \eqs{defamu}{defgammasat}.

To assess the stability of the medium, we then need to compare the change in buoyancy between a rising parcel and the environment. Assuming, as usual, that pressure equilibrates instantaneously, the criterion for convection becomes\footnote{\label{disc:condensate} This criterion assumes that the sedimentation of condensates is instantaneous so that their mass loading effect is negligible for both the environment and the rising parcel. As discussed in \citet{Gui95} and demonstrated in \app{app:moistconvection}, this is actually the most favorable case for convection because condensate are more abundant in rising parcels where condensation occurs \citep{WH06}. Mass loading by condensates is thus always an impediment to convection, even in a more familiar Earth-like atmosphere. This will be discussed at length in \sect{sec:limitations}. }
\balign{\left(\delt-\delmu\right)_\mathrm{env}-\left(\delt-\delmu\right)_\mathrm{parcel}>0.}
Using \eq{delmudelt}, and assuming that the parcel follows a moist adiabat, it is straightforward to show that this translates into
\balign{\left(\delt-\delmoist\right)\left(\amu\,\gammasat-1\right)>0,}
as was already found by \citet{Gui95}. An important difference with the \textit{dry} case is that the \mmw effect does not come as an additive factor, but as a multiplicative one. It thus acts as a conditional criterion on the amount of vapor present that can stabilize an otherwise unstable medium. Indeed, even if the thermal gradient is super-moist-adiabatic, convection is inhibited if
\balign{\label{stabcriterion}
\amu \gammasat>1 \Leftrightarrow \mratio \qvap \dd{\ln \psat}{\ln T}>1\Leftrightarrow\qvap \left(1-\frac{\Mair}{\Mvap}\right) \frac{\Mvap\latent}{\Rgp \temp} >1.}
In other words, moist convection is inhibited if the mass mixing ratio of vapor exceeds a critical mixing ratio
\balign{\label{qcri}
\qcri\equiv\frac{1}{\mratio} \frac{\Rgp \temp}{\Mvap\latent}.
}

\subsection{The cloud sequence}

This critical mixing ratio can be used to measure the potential impact of a given species. In an attempt to identify the most important one, we will follow the sequence of clouds that we expect in order of increasing condensation temperature. The results are summarized in \tab{tab:species}\footnote{The enrichments computed here use recent estimates of the solar oxygen and nitrogen abundances recommended by \citet{Lod10}, so that comparison with enrichment factors from the literature must be done with care, even if the mass mixing ratios are the same.}.

\begin{table}
\centering
\begin{tabular}{lrrrr}
\hline
& \textit{\textbf{\ch4}}	& \textit{\textbf{\nh3}} & \textit{\textbf{\h2o}} & \textit{\textbf{Fe}}\\
\hline
$\mratio$ & 0.85 & 0.86 & 0.87 & 0.96 \\
$\temp_\mathrm{ref} (K)$ & 80 & 150 & 300 & 3500 \\
$\Mvap \latent/\Rgp \temp_\mathrm{ref}$ & 12. & 19. & 16. & 12.\\
Critical mixing ratio ($\qcri$)& 0.10 & 0.062 & 0.070 & 0.089\\
\multicolumn{1}{l}{Enrichment over solar} & 40. & 78. & 9.9 & 74.\\
\hline
\end{tabular}
\caption{Critical mass mixing ratio over which moist convection is inhibited for four of the most abundant condensible species. The conversion into enrichment factor over solar uses recent solar abundances estimates recommended by \citet{Lod10}, which explains the differences with the values derived in \citet{Gui95}. $\temp_\mathrm{ref}$ is a crude estimate of the temperature at which the cloud deck of a given species would form in an atmosphere similar to Saturn's but with a solar metallicity (note that methane does not condense in Saturn, and the temperature is estimated for Uranus)}.
\label{tab:species}
\end{table}


\textbf{\ch4}:
Methane does not condense in Jupiter and Saturn, but there is evidence for such stabilized layers where \ch4 condenses (around 1-2\,bar) in Uranus and Neptune \citep{Gui95}. In these planets, the enrichment in \ch4 is about 80 times the solar value \citep[][and references therein]{Guillot+Gautier2015}, well above the critical value (see \tab{tab:species}). The effect is, however, rather small because at this depth, radiation is still an efficient energy carrier. Furthermore, our ability to probe temperatures below this region entails that the effect of this superadiabaticity has already been implicitly taken into account in the various interior modeling of these planets. 

\textbf{\nh3}:
Between 0.5 and 2 bars, ammonia does not seem to hamper convection in Jupiter and Saturn. This is to be expected because the mass mixing ratio of vapor needs to be relatively high below the cloud deck for condensation to have a significant effect (6\% in mass). Hence, nitrogen being much less cosmically abundant than oxygen or carbon, a huge enrichment, $\sim80$ times solar would be needed. The measured enrichments in ammonia in Jupiter \citep{NAC96,WMA04} and Saturn \citep{Fletcher+2011} are more than an order of magnitude smaller. 

In Uranus and Neptune, it is conceivable that the C/N ratio is solar and that ammonia is also super-critical, but this remains speculative. Owing to its condensation at deep levels, we have no spectroscopic constraints on the abundance of ammonia or on the temperature profile in these regions \citep[e.g.,][]{Guillot+Gautier2015}. 

\textbf{\h2o}:
Oxygen being the most abundant atomic species after hydrogen and helium, an O abundance greater than $\sim$10 times solar would be sufficient to stop convection in all four giant planets \citep{Gui95}. While this is close to the inferred value for Jupiter and Saturn, this is almost an order of magnitude smaller than the expected enrichment for the icy giants (assuming that the C/O ratio remains close to solar in these planets).

Heavier, more refractory species such as silicates, iron, and aluminium oxides have both a smaller expected abundance and a high cloud-deck temperature. As a result, water is our best candidate for the formation of a very superadiabatic, stable layer inside any cool, significantly enriched giant planet. We will thus focus this study on the effect of that species. 

\subsection{Which criterion to use?}\label{sec:whatcriterion}

\begin{table}
\centering
\begin{tabular}{lcccc}
\hline
& $\qvap= \qsat$& $\qvap \geqslant\qcri$ & Moist  & Dry\\
&	& &conv. &conv.\\
\hline
$\delt<\delmoist$ & &  & \xmark & \xmark \\
\hline
$\delmoist$ & \xmark &  & \xmark & \xmark \\
\hspace{0.3cm}$<\delt<$ & \cmark & \xmark & \cmark & \xmark \\
\hspace{0.8cm}$\delad+\delmu$ & \cmark & \cmark & \xmark & \xmark \\
\hline
$0<\deltetav$ & \xmark &  & \xmark & \cmark \\
\hspace{0.3cm} $\Leftrightarrow$& \cmark & \xmark & \cmark & \cmark \\
\hspace{0.2cm}$\delad+\delmu<\delt$& \cmark & \cmark & \xmark & \textcolor{red}{\cmark} \footnote{Although it would seem from applying the various criteria that dry convection should be allowed in this case, we put a red mark to remind the reader that this situation almost never occurs in practice, as discussed in \sect{sec:whatcriterion}.}\\
\hline
\end{tabular}
\caption{Summary of convective processes at play in various situations with $\delmu>0$. First, choose the temperature gradient regime on the left and whether the medium is saturated ($\qvap= \qsat$) and/or water rich (in the sense $\qvap \geqslant\qcri$). Then, the two right columns tell you which type of convection will develop. The absence of any symbol in a given box means that this specific criterion has no influence under the given conditions.}
\label{tab:criteria}
\end{table}

Finally, let us summarize the criteria to use in various situations. In most cases that we will encounter in planetary atmospheres, temperature and vapor mixing ratios decrease upward so that $\delmu>0$ and $\delmoist<\delad<\delad+\delmu.$ But it is not sufficient to compare thermal gradients to decide whether convection occurs or not. Indeed, for moist convection, the medium needs to be saturated ($\qvap=\qsat$) and below the critical vapor mixing ratio ($\qvap<\qcri$). We thus summarize all possible situations in \tab{tab:criteria}, which tells us when convective transport processes are efficient. This can be used to create a forward model of the atmosphere as done in \sect{sec:profiles}.

But let us focus a moment on what happens toward the bottom of the cloud deck of a fairly enriched atmosphere (in the sense that $\qvap>\qcri$). The presence of clouds means that we are near saturation. Looking at \tab{tab:criteria}, it would seem that the only possibility to have a significant convective transport is that the thermal gradient be Ledoux unstable. However, this situation does not arise in practice. Indeed, when the medium is saturated, the \mmw gradient is linked to the thermal gradient, and one can use \eqs{defamu}{delmudelt} to show that
\balign{\delt-\delmu-\delad=\delt\left(1-\amu\gammasat\right) -\delad+\mratio \qsat.}
Since we are in a region where $1-\amu\gammasat<0$ and where the thermal gradient must be super-adiabatic to transport the flux, this yields
\balign{\delt-\delmu-\delad<\mratio \qsat -\amu\gammasat\delad= \mratio \qsat\left(1-\frac{\Mvap\latent}{\Rgp\temp}\right)<0,}
in the range of temperatures around the cloud deck (see \tab{tab:species}).

As a result, overturning convection cannot transport the internal flux near the bottom of the cloud deck if the atmosphere is sufficiently enriched. Before proceeding on to the modeling of our giant planets, the question that remains to be elucidated is whether an other hydrodynamical instability arises and is able to turbulently carry this flux, or if only radiation is at play, creating a stable, strongly super-adiabatic radiative layer. This is the goal of the next section.

\section{Linear analysis of the double-diffusive instability in a condensable medium}\label{sec:linear_anlysis}

In a real atmosphere, both heat and vapor are allowed to diffuse, and with varying efficiency. This has been shown to cause various instabilities that can affect the properties of the energy transport in a otherwise convectively stable layer \citep{Ste60}. However, condensation is usually not accounted for in the study of these so-called double diffusive processes \citep{Ste79,RGT11}.
To address this shortcoming, in \sect{sec:boussinesq} we thus carry a linear analysis using the Boussinesq approximation \citep{Bou72}, but where we implement the effect of condensation (both latent heat and mean molecular weight effects). 

We will first show that this new set of equations can recover expected behavior in well known limit cases (\sect{sec:inhibition}). Then, we will demonstrate in \sect{sec:killingDDconv} that condensation can actually kill the double diffusive instability in a saturated medium, leaving radiation as the sole viable mean to carry the planetary flux when moist convection is inhibited. 

\subsection{Linear analysis}\label{sec:boussinesq}

\subsubsection{Basic equations}

We consider an infinite medium in a uniform gravity field $\vg=-\grav \,\uz$. The velocity, pressure, temperature and density of the gas are denoted by $\vv$, $\press$, $\temp$ and $\rho$. The equation of state has been described in \sect{sec:conv_adiabatic}.
The Boussinesq system for a two-phase fluid accounting for condensation writes
\balign{
&\nabla \cdot \vv=0,\\
&\rho\,\bigd{t} \vv=-\overrightarrow{\nabla }\press +\rho \,\vg+\rho \nu\nabla^2 \vv,\\
&\bigd{t} \temp-\frac{1}{\rho\,\cp}\bigd{t} \press=\kapt\nabla^2 \temp +\frac{\latent}{\cp}\frac{1}{1-\qvap}\frac{\qvap-\qsat}{\tcond}  \label{DtT}\\
&\bigd{t}\qvap =D\,\nabla^2 \qvap -(\qvap-\qsat)/\tcond, \label{Dtq}
}
where $D$, $\kapt$ and $\nu$ are the solute, thermal and kinematic viscosities (or diffusivities, in units of length squared over time). $\bigd{t}\equiv \ppd{t}+\vv\cdot \overrightarrow{\nabla}$ is the Lagrangian derivative operator. The effect of condensation is implemented by adding the last term in the energy equation, \eq{DtT}, which accounts for the latent heat release -- without implying that the vapor is a trace gas -- and the last term in \eq{Dtq}, which tends to restore saturation by either condensing vapor or vaporizing condensates. This introduces an important model parameter, $\tcond$, which represents the timescale on which condensation/sublimation restores saturation. It is in fact a very important parameter as it will serve to parametrize the efficiency of condensation and whether it is faster or slower than other processes affecting the vapor.

\subsubsection{The background state}

The mean field is characterized by a null velocity, a pressure gradient, $\overrightarrow{\nabla } \pressm =\rhom \,\vg$, which introduces a reference pressure scale height $\hp=\pressm/(\rhom \grav)$, and a temperature gradient, $\delt=-\hp \, \partial_z \ln\tempm.$
Mean quantities are identified by an overbar.
For future reference, we note that the dry adiabatic temperature gradient can be linked to the pressure gradient by
\balign{\delad =-\frac{\hp}{\tempm}\frac{1}{\rhom\,\cp} \partial_z \pressm=\frac{\Rgp}{ \cp\,\mum},}
and introduce the well known \brunt frequency $\Nt^2~\equiv~\frac{\grav}{H}(\delt-\delad).$

Finally, the mean field also exhibits a vapor mixing ratio gradient, $\delq\equiv  \d \ln \qvap/\d\ln \press=-\hp\,\partial_z \ln \qvapm=\gammavap\delt,$
with $\gammavap\equiv\left.\pd{\ln \qvap}{\ln T}\right|_\press,$ and the associated frequency 
$\Nq^2\equiv\frac{\grav}{\hp}\delq.$
The \mmw gradient is given by $\delmu=\amu \delq$.

In the framework of a linear analysis, the mean state must be solution of the time independent equations. Thus, whenever condensation is allowed ($\tcond \neq \infty$), it is implicitly assumed that $\qvapm = \qsatm$ and that the water vapor gradient follows the saturation vapor curve ($\gammavap=\gammasat$). We will however keep differentiating these quantities to show that in the absence of condensation (where the vapor gradient is a free parameter), we recover the usual double-diffusive instability for an incondensable species.

\subsubsection{Linearized equations}

 The linearized Boussinesq equations around this state write
\balign{
&\nabla \cdot \dvv=0,&\\
&\left( \ppd{t} -\nu\nabla^2 \right)\dvv=\frac{\drho}{\rhom} \vg,&\\
&\left( \ppd{t} -\kapt\nabla^2 \right) \deltemp+\dvv\cdot \left( \overrightarrow{\nabla} \tempm- \left.\overrightarrow{\nabla} \tempm\right|_\mathrm{ad}\right)=\frac{\latent}{\cp}\frac{1}{1-\qvapm}\frac{\delqvap-\delqsat}{\tcond} ,\\
&\left( \ppd{t} -D\nabla^2\right)\delqvap+\dvv\cdot\overrightarrow{\nabla }\qvap= -(\delqvap-\delqsat)/\tcond,\label{lineardqvap}\\
&\frac{\drho}{\rhom}=  \frac{\dmu}{\mum} - \frac{\deltemp}{\tempm}=\amu \frac{\delqvap}{\qvapm} -\frac{\deltemp}{\tempm},\label{lineareos}
}
where all the perturbations to the mean field have a $\delta$. Notice that because the temperature is perturbed, the saturation mixing ratio at the given pressure level is also perturbed following the Clausius-Clapeyron law\footnote{Pressure perturbations are disregarded in our approximation, so that $\delqvap$ does not have any pressure term.}, 
\balign{
&\frac{\delqsat}{\qsatm}=\left.\dd{\ln \qsat}{\ln T}\right|_\press\frac{\deltemp}{\tempm}\equiv \gammasat\frac{\deltemp}{\tempm}.
}

An important assumption made here is that condensates are instantaneously removed from a rising/cooling parcel where condensation occurs, which explains why there is no condensed phase term in the linearized equation of state\footnote{One could actually argue that if condensates are always retained during condensation, density should always increase when the vapor amount decreases (the total mass does not change but the total volume decreases), contrary to what \eq{lineareos} seems to imply. This is not so. Indeed, in our Eulerian framework, local vapor variations are also due to diffusion and advection, and are not necessarily linked to a change in the amount of condensates.} (\eq{lineareos}).
At the same time, we also assume that a (small) quantity of condensates is always available to be vaporized if the fluid is subsaturated, generally on descent. These are usual assumptions made when computing a moist adiabat. Although possibly stringent, it has to be kept in mind that these assumptions are needed to keep the system linear. As demonstrated in \app{app:moistconvection}, however, including these effect could potentially suppress convective motion even more efficiently. 
These points will be discussed in detail in \sect{sec:limitations}.

\subsubsection{Dispersion relation for plane waves}

We want to know the response to  plane waves of dependency $\propto e^{\freq t+i \vk \cdot \vr}.$ Thus, the continuity equation gives us $\vk\cdot \dvv=0$ and we know that the wave vector must be perpendicular to the velocity perturbation. Because the gravitational forcing is vertical, it can be shown that the most unstable mode will always have a vertical velocity (elevator mode), and we can project all the equations along this axis without loss of generality \citep{RGT11}.

With the notation above, the equations for the conservation of momentum, energy, and vapor become
\balign{\label{linearmomentum}
\left(\freq +\nu \,\wn^2 \right)\dv=-\grav\frac{\drho}{\rhom} =-\grav\left(\amu \frac{\delqvap}{\qvapm} -\frac{\deltemp}{\tempm}\right),
}
\balign{\left(\freq +\kapt\,\wn^2 +\gammasat\frac{\qsatm}{\qvapm}\frac{\betacond}{\tcond}\right) \frac{\deltemp}{\tempm}&=\left(\delt-\delad\right)\,\frac{\dv}{\hp}+\frac{\betacond}{\tcond}\,\frac{\delqvap}{\qvapm} \nonumber\\
&= \Nt^2 \,\frac{\dv}{\grav}+\frac{\betacond}{\tcond}\frac{\delqvap }{\qvapm},
\label{linearenergy}
}
\balign{
\left(\freq +D\,\wn^2 +\itcond\right)\frac{\delqvap}{\qvapm}&= \frac{\delmu}{\amu\hp}\dv+ \frac{\delqsat}{\tcond\,\qvapm}\nonumber\\
&=\frac{\Nmu^2}{\amu}\frac{\dv}{\grav}+\frac{ \qsatm}{\qvapm} \frac{\gammasat}{\tcond}\frac{\deltemp}{\tempm},
\label{linearqvap}
}
where we define a latent heat parameter
\balign{\label{betacond}
\betacond\equiv\frac{\latent}{\cp T}\frac{\qvapm}{1-\qvapm}.}

To attain the dispersion relation, our first goal is to express $\deltemp$ and $\delqvap$ as a function of $\dv$ only using \eqs{linearenergy}{linearqvap}. 
This yields
\balign{\label{deltatempfinal}
&\freqK\frac{ \deltemp}{\tempm}=\left(\Nt^2+ \frac{\betacond}{\freqD\tcond}\Nq^2\right) \,\frac{\dv}{\grav},\\
&\label{deltaqvapfinal}
\freqD\frac{ \delqvap}{\qvapm}=\left[\frac{\qsatm}{\qvapm}\frac{\gammasat }{\freqK\tcond}\left(\Nt^2+ \frac{\betacond}{\freqD\tcond}\Nq^2\right)+\Nq^2\right] \,\frac{\dv}{\grav},
}
where, for compactness, we have defined three frequencies
\balign{
&\freqnu\equiv \freq +\nu\,\wn^2,\\
&\freqD\equiv \freq +D\,\wn^2 +\itcond,\\
&\freqK\equiv \freq +\kapt\,\wn^2 +\gammasat\frac{\qsatm}{\qvapm}\frac{\betacond}{\tcond}(1-\frac{1}{\freqD\tcond}).
}

Finally, by introducing these expressions into \eq{linearmomentum}, the velocity perturbation amplitude vanishes from the equations and we get the dispersion relation 
\balign{\label{generaldispersion}
&\freqnu=\frac{1}{\freqK}\left(\Nt^2+ \frac{\betacond}{\freqD\tcond}\Nq^2\right)\left( 1-\amu\frac{\qsatm}{\qvapm}\frac{\gammasat }{\freqD\tcond} \right)-\amu\frac{\Nq^2}{\freqD} .  }

Although rather complex, this dispersion relation has the advantage of being quite general. Then, it is very simple to retrieve the dispersion relation in well known limit cases. For example, one can turn off the effect of the \mmw by putting $\amu=0$, the latent heating by putting $\betacond=0$, or directly the condensation altogether by putting $\tcond\rightarrow\infty$. The regular adiabatic limit can also be found by putting all the diffusivities to zero.

\subsection{Application to well-known limit cases}\label{sec:inhibition}

Here we will first show that this set of equation recovers well known behavior in several limit cases:
\begin{itemize}
\item[$\bullet$] \sect{sec:ledoux}: The Schwarzschild and Ledoux criteria in the dry, adiabatic limit.
\item[$\bullet$] \sect{sec:DDinstab}: The double diffusive instability in the dry regime.
\item[$\bullet$] \sect{sec:adiablimit}: Moist convection and its inhibition in the adiabatic limit, confirming the simpler derivation made in \sect{sec:conv_adiabatic_mmw}.
\end{itemize}
Then, in \sect{sec:killingDDconv}, we will show that if condensation is occurring on a much shorter timescale than vapor diffusion, the medium is stable against the double-diffusive instability.

\subsubsection{The dry, adiabatic limit: Schwarzschild and Ledoux criteria}\label{sec:ledoux}

To start simple, let us consider the dry ($\tcond\rightarrow\infty$), adiabatic ($\kapt=\nu=D=0$) limit. In this regime, $\freqnu=\freqK=\freqD=\freq$, and \eq{generaldispersion} straightforwardly yields
\balign{
&\freq^2=\Nt^2-\amu\Nq^2=\frac{\grav}{\hp}\left(\delt-\delad-\delmu\right).
}
The instability will thus grow---i.e. $\freq>0$---if $\delt>\delad+\delmu$, which is the usual Ledoux criterion (or Schwarzschild's if there is no \mmw gradient).

\subsubsection{The double-diffusive instability in the dry regime}\label{sec:DDinstab}

Staying in the dry limit, but allowing for some diffusion, \eq{generaldispersion} becomes
\balign{ \label{doublediff_dispersion}
&(\freq +\nu\,\wn^2)(\freq +D\,\wn^2)(\freq +\kapt\,\wn^2)=\nonumber\\
&\ \ \ \ \ \ \ \ \ \ \ \ \ \ \ \ \ \ \ \ \ \ \ \ \ \ \ \ \ \ \ \ \ \ \ \Nt^2\,(\freq +D\,\wn^2) -\amu\Nq^2\,(\freq +\kapt\,\wn^2) .
}
We recover the usual equation for double-diffusive convection \citep{Ste60}. Notice that the usual inverse density ratio used in the study of double diffusive convection takes into account the effect of the mean molecular weight and is $\Rrho^{-1}=\amu(\Nq/\Nt)^2$ \citep{RGT11}.
 The medium is thus unstable when
\balign{1\leq \amu\frac{\delq}{\delt-\delad} \leq \frac{1+\Pr}{\taud+\Pr},}
$\Pr\equiv\nu/\kapt$ is the usual Prandlt number, and $\taud\equiv D/\kapt$ the diffusivity ratio.
This shows that the double-diffusive instability is captured by our set of equations.

\begin{figure}[htbp] 
 \centering
\subfigure{ \includegraphics[scale=.56,trim = 0cm .cm 0.cm 0.cm, clip]{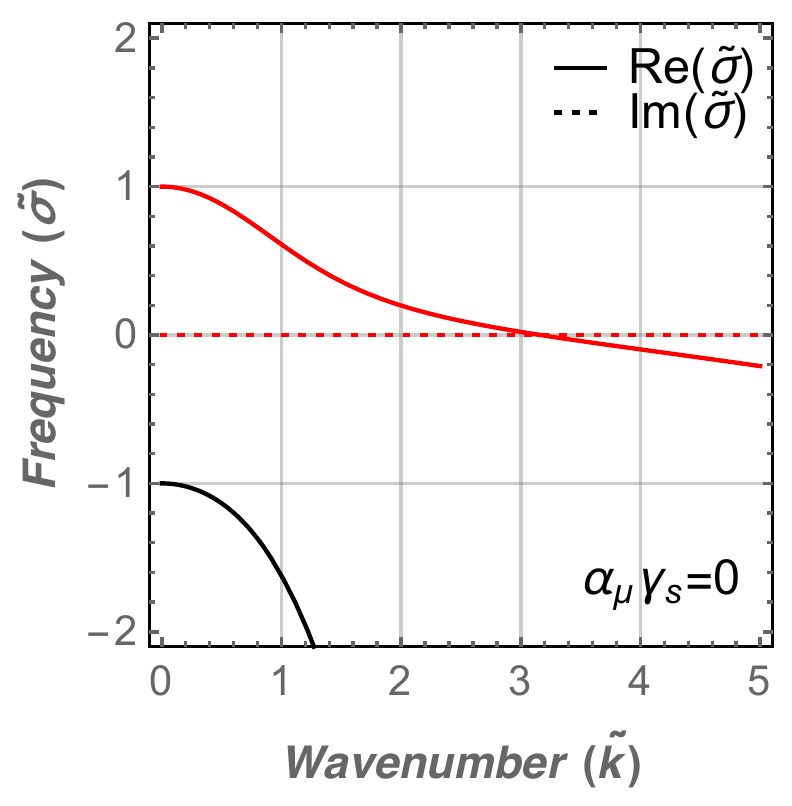} }
\subfigure{ \includegraphics[scale=.56,trim = 1.4cm .cm 0.cm 0.cm, clip]{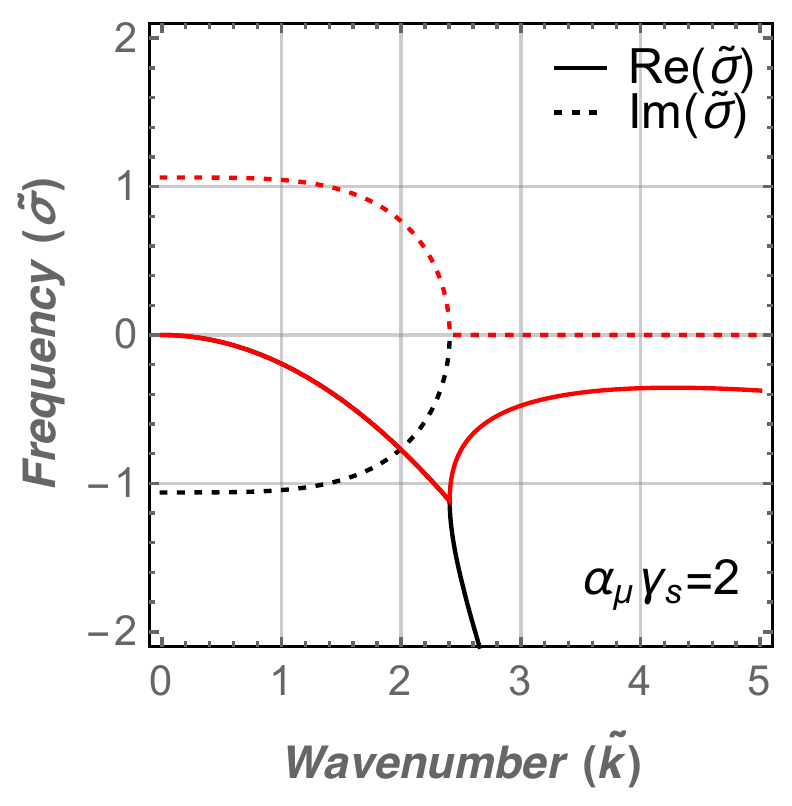} }
\caption{
Growth rate of the two modes given by \eq{efficientcond}. The real part is denoted by a solid curve (positive means a growing mode, negative means a damped mode) and the complex part (frequency of the oscillating mode) by a dotted line. Colors are used to distinguish the two modes, red showing the most unstable one. The left panel corresponds to an unstable case for which criterion \ref{stabcriterion} is not met ($\qvap\rightarrow 0\Rightarrow\amu=\betacond=0$). The right panel shows a case where the effect of the mean molecular weight is sufficient to stabilize the medium ($\amu\gammasat=2,\ \betacond\gammasat=1.7$). Only a damped oscillating mode remains. Other umerical values are $\Pr=\taud=0.01$ and $(\Nq/\Nt)^2=30$ (see \sect{sec:adiablimit}).
}
 \label{fig:linstab}
\end{figure}

\subsubsection{Moist convection}\label{sec:adiablimit}
If we now allow for an efficient condensation ($\tcond\rightarrow0$, i.e. $\tcond$ is smaller than every other timescale in the problem), and turn off diffusion ($\kapt=\nu=D=0$), we should recover the behavior of adiabatic moist convection discussed earlier. Indeed, in this case, $\qvapm=\qsatm$ and 
\balign{\freqnu=i\freq,\ \ \freqD=\itcond,\ \ \freqK=\freq\,(1+\betacond\gammasat).}
 As a result, the general dispersion relation, \eq{generaldispersion}, simplifies to
\balign{
\freq^2 =\left(1-\amu \gammasat  \right) \left(\Nt^2+\betacond\Nq^2\right)/(1+\betacond\gammasat). 
}
The medium is stable if $\freq^2<0$ which happens if only one of the quantities in parentheses is negative. Using \eq{delmudelt}, we can rewrite the condition on the second parenthesis as follows
\balign{\delt<\delad/(1+\betacond\gammasat)\equiv \delmoist,}
which is the usual condition that a saturated medium is stable if the thermal gradient is less than the moist adiabat\footnote{Comparing this expression to \eq{moistadiabat}, one can see that a term is missing. Upon further inspection, this term is in fact a compression term that is implicitly dropped when the Boussinesq approximation is made.}. As we assume that our medium would be unstable in the absence of any \mmw effect, this stability criterion in never verified. The medium is thus stable only if $\amu \gammasat>1$, as advertised.

Let us give some numerical reference values. For water condensation in hydrogen around 300\,K, $\gammasat\approx 16.$ The threshold for stability is thus around $\qcri=0.06-0.07.$ In \fig{fig:linstab} we show a case with $\amu\gammasat= 2,$ which corresponds roughly to $\qvapm\approx0.15,$ $\amu\approx0.15,$ $\betacond\approx0.09,$ and $\betacond\gammasat\approx 1.2.$ Finally, one needs $(\Nq/\Nt)^2=\delq/(\delt-\delad)$ which can be constrained considering that $\delq/\delt\approx \gammasat$ and $\delt/(\delt-\delad)>1.$ We will therefore always use $(\Nq/\Nt)^2 > \gammasat$. In \fig{fig:linstab}, we use the fiducial value of 30.

\subsection{Efficient condensation limit, or how to kill an instability}\label{sec:killingDDconv}

Now, we only assume rapid condensation while allowing for diffusion. This means that $\tcond$ is the shortest timescale in our problem. Then, taking the dispersion relation, \eq{generaldispersion}, in the $\ttcond\equiv\tcond\Nt\rightarrow0$ limit, we get
\balign{\label{efficientcond}
&(\tfreq +\Pr\,\tk^2)\left(\tfreq\, (1+\betacond\gammasat)+(1+\betacond\gammasat \taud)\,\tk^2\right)=\nonumber\\
&\ \ \ \ \ \ \ \ \ \ \ \ \ \ \ \ \ \ \ \ \ \ \ \ \ \ \ \ \ \ \ \ \ \ \ \ \ \ \ \ \ \ \ \ \ \ \ \ \ \ \ \ \left(1-\amu \gammasat  \right) \left(1+\betacond(\Nq/\Nt)^2\right),
}
where we have used the inverse \brunt frequency as our unit of time and the thermal lenghtscale ($\sqrt{\kapt/\Nt}$) as a unit of length, defining $\tfreq\equiv\freq/\Nt$ and $\tk\equiv \sqrt{\kapt/\Nt}\,\wn$.


For any given wavenumber, this equation of degree two has two complex solutions, say $\tfreq_1$ and $\tfreq_2$, defining two possible modes.
These modes can be computed analytically, and are shown in \fig{fig:linstab} for two cases, i.e. with and without condensation-inhibited convection.

Actually, even without specifying any parameter, we can demonstrate analytically that the aforementioned adiabatic stability criterion ($\amu \gammasat>1$) remains unchanged in the efficient condensation regime with diffusion. Indeed, \eq{efficientcond} is of the form $\tfreq^2+b\tfreq+c=0$ with both $b=-(\tfreq_1+\tfreq_2)$ and $c=\tfreq_1\tfreq_2$ real and positive. This imposes that $\Im(\tfreq_1)=-\Im(\tfreq_2),$ and, consequently, $\Re(\tfreq_1)\Re(\tfreq_2)\ge0$, so that the real parts of the two solutions (the two growth rates) must have the same sign. Finally, because $b=-\Re(\tfreq_1+\tfreq_2)\ge0$, we can infer that the two growth rates are negative. 

 This means that, whenever $\amu \gammasat>1$ and condensation is efficient, the double diffusive instability is killed, and there is no growing, overstable mode, independently of the various diffusivity ratios.

Where does this come from?
The physical reason is rather simple. If condensation is efficient, the amount of water vapor at any given level is set by thermodynamics, and thus, by the temperature at that level. Thus, any mechanism affecting the temperature field, such as thermal diffusion, effectively affects the water vapor field as well. In other words, water vapor effectively "diffuses" as fast as heat, hence the absence of "double-diffusive" processes which requires that one of the components, generally the solute, diffuses much slower than the other. 

\subsection{Dimensionless equation and numerical limits}

\begin{figure}[tbp] 
 
\subfigure{ \includegraphics[scale=.7,trim = 0cm .cm 0.cm 0.cm, clip]{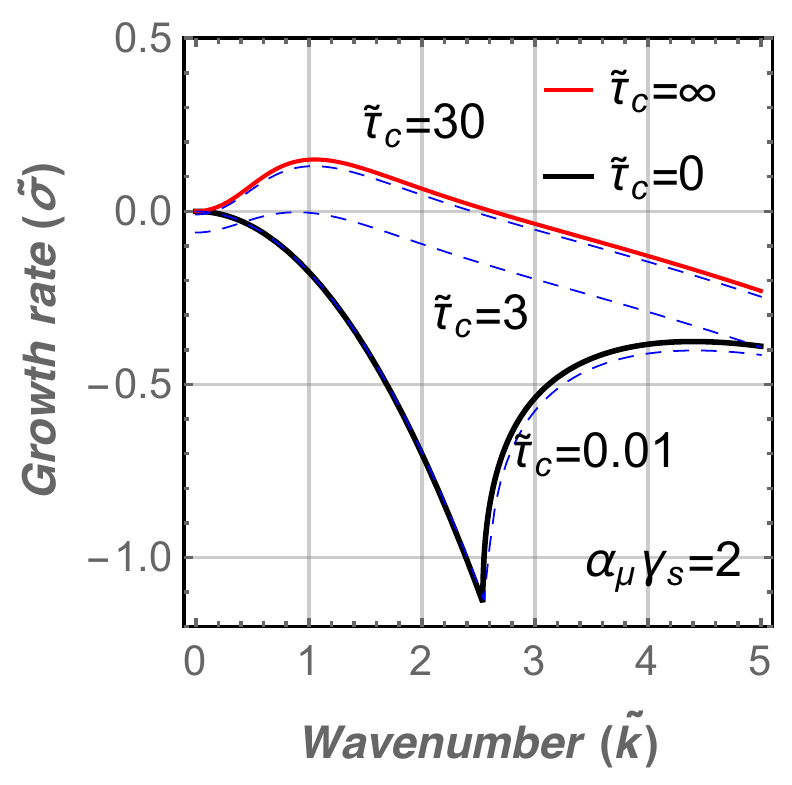} }
\caption{
Growth rate of the fastest growing mode of the general dispersion relation. The solid curves represent the limiting case of efficient condensation (black; \eq{efficientcond}) and no condensation (red; \eq{doublediff_dispersion}). From top to bottom, the three dashed blue curves are for $\ttcond=$30, 3 and 0.01 respectively.
Other Numerical values are $\amu\gammasat=2,\ \betacond\gammasat=1.7$, $\Pr=\taud=0.01$ and $(\Nq/\Nt)^2=30$. When $\ttcond$ decreases, the medium undergoes a transition from a regime where the double-diffusive mode can grow (positive growth rate) to a regime where this instability is shut-down by condensation.
}
 \label{fig:linstabtau}
\end{figure}

To show that the result derived above is still valid for a finite condensation time, we solve the dispersion equation numerically. But we deal only with dimensionless number by using $\Nt^{-1}$ as a timescale and $\sqrt{\kapt/\Nt}$ as our lengthscale. For simplicity, we also assume that the mean state is in condensation equilibrium ($\qvapm=\qsatm$). This yields
\balign{\label{dimensionless_dispersion}
&\freqnut\freqKt\freqDt+\freqKt\amu\tNq^2 =\freqDt\left(1+ \betacond\frac{\tNq^2 }{\freqDt\ttcond}\right)\left( 1-\amu\frac{\gammasat }{\freqDt\ttcond} \right),  }
where $\tNq\equiv\Nq/\Nt$, and
\balign{
&\freqnut\equiv \tfreq +\Pr\,\tk^2,\\
&\freqDt\equiv \tfreq +\taud\,\tk^2 +\titcond,\\
&\freqKt\equiv \tfreq +\tk^2 +\gammasat\frac{\betacond}{\ttcond}(1-\frac{1}{\freqDt\ttcond}).}

This equation can be solved numerically. In \fig{fig:linstabtau}, we show the growth rate of the fastest growing mode that is solution of \eq{dimensionless_dispersion}. One can see in the figure that the full solution smoothly recovers the various limiting case when $\ttcond$ is varied.

This figure further shows that, when realistic numerical values are put in, $\ttcond\lesssim3$ is sufficient to kill the double diffusive instability. This number reduces only by a factor of two when the Prandlt and diffusivity ratio are decreased by two orders of magnitude, or when $\amu \gammasat$ is decreased to $\approx 1.01$ ($\qvap$ close to the critical threshold). Increasing $\amu \gammasat$, the Prandlt, and/or the diffusivity ratio  actually relaxes the constraints on $\ttcond$ which can be even higher. So the timescale for cloud particle growth by condensation needs to be smaller than a few times the timescale for the overturning of an eddy which can be in excess of several hours in giant planet atmospheres. This seems reasonable. Indeed, for giant planet environments, \citet{Ros78} found that the full development of clouds and precipitations would occur in less than 10$^3$\,s. 

\subsection{Limitations of the linear analysis}\label{sec:limitations}

\subsubsection{Up/Down-draft asymmetry and linearity}

In our analysis, we basically assume that the medium is always saturated and that condensates are efficiently removed or resupplied if need be. 

It should be clear that, because condensation is strongly linked to updrafts and subsaturation to downdrafts \citep{WH06}, moist processes create a strong asymmetry in the system to be studied. Starting from a clear atmosphere, one would have to enforce different equations of motion for rising or sinking eddies. This would prevent any attempt at a linear analysis. 

How our assumptions on condensates affect double-diffusive convection is, however, not trivial to assess. Hereafter, we thus discuss how both condensates retention and subsaturation tend to suppress any rising/sinking motion, respectively. This seems to lend support to the idea that our criterion is conservative. In other words, accounting for these additional effects would suppress double-diffusive processes even more efficiently. This conclusion is however highly tentative and awaits confirmation by further experiments. 

\subsubsection{Effect of mass loading by condensates}

The first side of our approximation is that condensates are instantaneously removed from a condensing, generally ascending, parcel. This is common approximation made when computing, for example, the moist adiabat. But actually, it can be shown that \textit{any amount of condensates retained during ascent tends to hamper the rising motion} (see \app{app:moistconvection}). Condensates forming during adiabatic cooling always weigh down on updrafts, as aknowledged on Earth \citep{WH06}.


How does this mass loading affect the onset of \textit{double-diffusive} convection is a subtle question. We will have to wait for numerical or laboratory experiments to fully answer it. For the moment, it seems reasonable to argue that if the retention of condensates is an impediment to large-scale convection -- and in fact to any rising motion -- it should also be an impediment to small-scale double-diffusive convection.

\subsubsection{Stability of dry, subsaturated downdrafts}

On the other side, we also assume in our linear framework that when an eddy sinks (downdraft) condensates are available to keep the gas saturated. 
But because condensates are much denser than the gas, these two phases can often decouple---condensed particles leaving the eddy on ascent. In general, this causes subsident regions and downdrafts to be relatively dry and subsaturated \citep{WH06}.

Although not accounted for, we expect that subsaturation will only enhance the stabilization effect of condensation discussed here.
Indeed, consider a sinking eddy in a super-moist-adiabatic region where convection is inhibited by condensation, as described above. In our linear picture, the stabilization comes from the fact that the mean molecular weight will increase less rapidly in the eddy than in the environment. The buoyancy of the eddy is thus positive. Now, consider the same eddy but from which all the condensates are removed. On descent, the mean molecular weight in the parcel will not change at all because there is nothing to sublimate. The stabilizing effect of the \mmw gradient in the environment is now maximum as it is not offset whatsoever by sublimation in the eddy (see \app{app:moistconvection}). 
In this limiting case we recover the usual \citet{Led47} stability criterion, and the argument discussed in \sect{sec:whatcriterion} can be used. So the criterion for the inhibition of moist convection by condensation always entails the stability of dry sinking eddies in a saturated environment.

\subsubsection{Unidimensional approach}

Our analysis is of course limited by its 1D character. In 3D, one could try to imagine scenarios where the structure of the atmosphere described here, two convective layers separated by a stable, diffusive interface, would be broken by strong, localized up/downward motion that would penetrate this interface. One could also imagine the development of non purely vertical modes.

While we cannot preclude the existence of such episodic event (that could be reminiscent of observed giant storms), let us stress that in cool atmospheres where volatiles condense, the enrichment \textit{must} be higher at depth than aloft, creating a composition gradient localized around cloud levels. It is thus difficult to go around the fact that in supersolar H/He atmospheres, composition will, on average, stabilize the atmosphere to some extent. 
 This has been confirmed by mesoscale 2D numerical simulations of moist convection for Jupiter \citep{NTI00,SON06}.

However, because of their coarse resolution, these simulations could not have captured any double-diffusive instability (even if diffusion had been included; \citealt{RGT11}). Our picture of a highly stable diffusive interface at the cloud level thus remains to be validated by experiments, numerical or otherwise, investigating double-diffusive processes in a condensable gas. 
 

\section{Thermal profiles for the atmosphere of Solar System giant planets}\label{sec:profiles}

In the previous sections, we have shown that under some conditions that could reasonably be met in giant planets, convective transport is inhibited. It is now time to quantify the implications of such an inefficient transport on the thermal structure of the atmosphere of our giant planets. 

\begin{figure}[htbp] 
 \centering
\subfigure{ \includegraphics[scale=.25,trim = .cm 1.cm .cm 1.cm, clip]{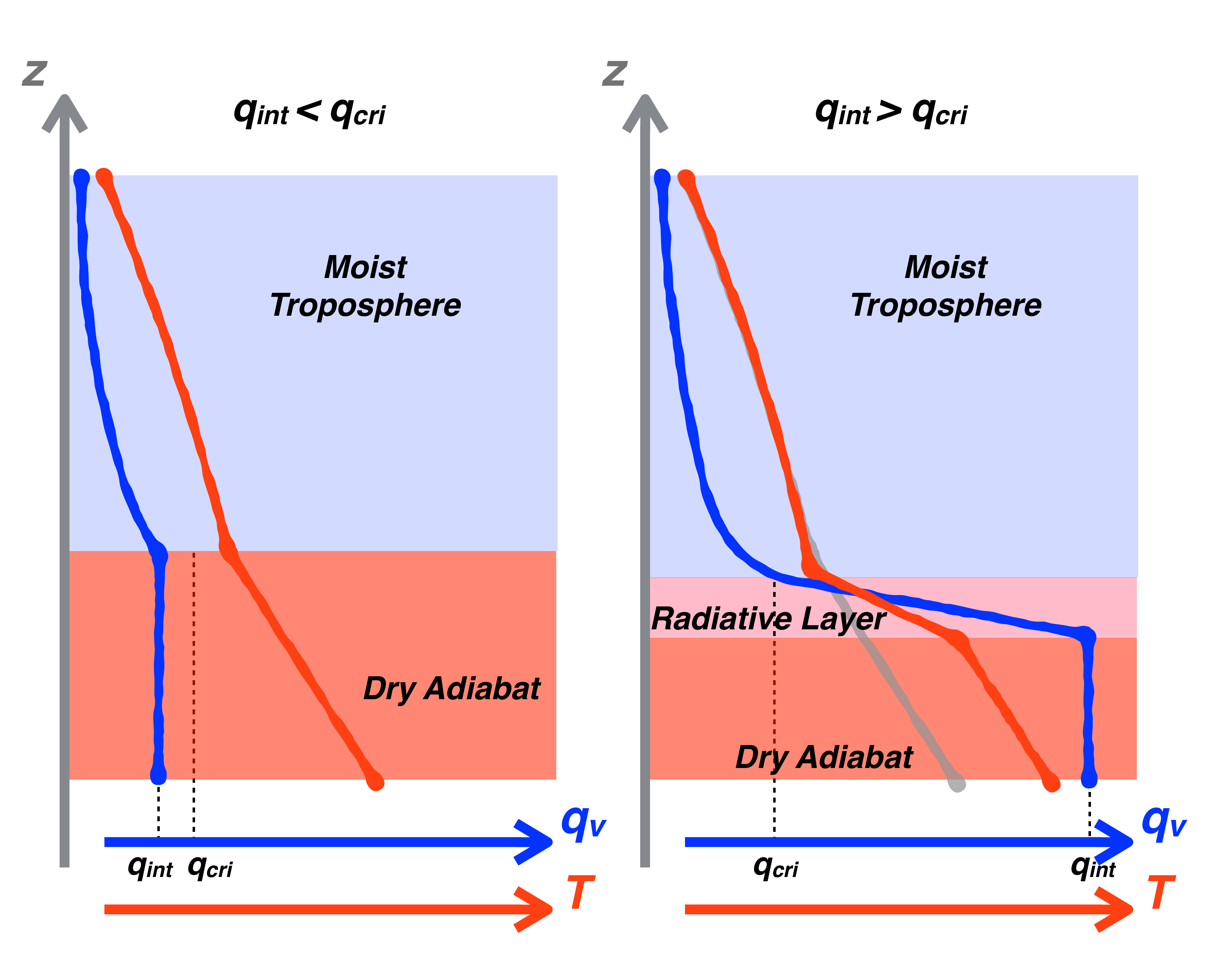} }
\caption{
Schematic of the layers present in the atmosphere. Blue and red curves depict the water mixing ratio and temperature profiles, respectively. \textit{Left panel:} Standard, 2-layer structure where a moist troposphere is underlain by a dry convective region ($\qint<\qcri$). \textit{Right panel:} When $\qint>\qcri$, convection is inhibited above the cloud deck and a third, radiative layer appears. A grey curve replicates the temperature profile of the 2-layer case to highlight the temperature increase in the deep adiabat.}
 \label{fig:3layers}
\end{figure}

\subsection{Numerical model}

To compute the atmospheric profiles, we assume that the atmosphere reaches a steady-state in hydrostatic and thermal equilibrium. Working in pressure coordinates, the goal is thus to integrate the temperature and vapor profiles downward from a given boundary condition. In practice, this model-top is chosen below the radiative-convective boundary, toward the bottom of the region probed observationally for each planet (see table~\ref{tab:planets}).

The most important aspect of the model is the choice of the prescription for the energy transport at a given level and the thermal gradient that results ($\delt$). We simply consider that the vapor is always brought to saturation when possible. Based on the analysis performed in \sect{sec:conv_adiabatic} and summarized in \tab{tab:criteria}, we envision a 3-layer structure shown in \fig{fig:3layers} with, from top to bottom,
\begin{itemize}
\item[$\bullet$] \textit{A moist, tropospheric layer}, where the internal flux is carried mostly by usual moist convection and where $\delt=\delmoist$. The vapor mixing ratio is equal to the saturation value. 
\item[$\bullet$] \textit{A stable, radiative layer}. If, and when the vapor mixing ratio reaches the critical value\footnote{Notice that, as $\qcri$ depends on the temperature, it is not a constant parameter throughout the atmosphere. In practice, the existence of a radiative layer is inferred by evaluating the criterion given by \eq{stabcriterion} at each pressure level.} ($\qcri$), we have shown that both moist and double-diffusive convections are inhibited. The energy is thus carried out by radiation and the temperature follows the radiative gradient ($\delrad$; see below). The vapor is still at saturation.
\item[$\bullet$] \textit{A deep adiabatic layer}. As the temperature keeps increasing, the vapor mixing ratio will reach the prescribed value for the deep interior, $\qint$. This is the bottom of the cloud deck. Below that point, dry convection is allowed, and turbulent exchange with the deep interior will homogenize the vapor mixing ratio without saturating the atmosphere. The atmosphere follows a dry adiabat, $\delt=\delad$.
\end{itemize}
The internal water mixing ratio, $\qint$, is our only free parameter. Of course, one can directly see that if $\qint<\qcri$, the conditions needed to have a stably stratified layer are never met, and we recover the usual 2-layer atmosphere with a moist troposphere underlain by a dry one. 

The radiative gradient is given by 
\balign{\delrad\equiv\frac{3}{16} \frac{\press\,\kapR  }{\grav} \frac{\Fint}{\ssb\temp^4},
}
where $\ssb$ is the Stefan-Boltzmann constant, $\kapR$ is the Rosseland mean opacity, and $\Fint$ is the internal cooling flux of the planet. The assumptions here are twofold. First, all the sunlight is assumed to be absorbed above the radiative layer, so that only the internal flux needs to be carried there. Second, the radiative layer must be deep enough for radiative transport to proceed in the diffusive limit. The parametrization for the Rosseland mean opacities are taken from \citet{VGP13}. This parametrization requires a metallicity, scaled on the elemental abundance of oxygen. The last approximation has however almost no consequence because the radiative gradient is always much larger than the adiabatic one in the regions where we expect the radiative layer to be. The radiative layer hence acts almost like a temperature jump whose magnitude is determined by water thermodynamics on one side and internal water content on the other. 

The specific heat capacity of water is taken from NIST and includes a temperature dependency. The saturation vapor pressure curve for water is computed using Tetens formula, 
$\psat=\press_\mathrm{1}\exp\left(b \frac{T-T_\mathrm{1}}{T-T_\mathrm{2}}\right)$ where $p_1=611.14$\,Pa, $T_1=273.16$\,K, and 
\balign{(T_2, b)=\left\{\begin{array}{lrllcc}(&35.86\,K, & 17.269 &)&  & T>T_1 \\(&7.66\,K, & 21.875&) &  & T<T_1\end{array}\right. .} The latent heat is derived using the Clausius-Clapeyron relation. The heat capacity of H$_2$ comes from \citet{VTC14} and can be calculated either for a normal or equilibrium mixture of ortho and para spin isomers. For Jupiter and Saturn, the deepest measured temperature is high enough for the two assumptions to yield the same results. For Uranus and Neptune, data do not go deep enough for this to be completely true and the temperature predicted at depth may somewhat depend on the assumption used. For simplicity, we will use the usual \textit{normal} ratio approximation. In the pressure-temperature domain of our study, we checked that there is no significant pressure effect on the heat capacity using pressure-dependent data from \citet{MHR81}. The specific heat capacity of He is assumed constant with $\temp$ and set to $5\Rgp/2 M_\mathrm{He}$. The total heat capacity is computed using the additive volume law.

The model top temperature ($\temp_\mathrm{top}$) and pressure ($\press_\mathrm{top}$), the internal flux, and He volume mixing ratio\footnote{These He mixing ratios are measured in the upper part of the atmosphere where all molecules except H$_2$ and He are in trace amounts. We thus assume that $\xHetop+x^\mathrm{top}_\mathrm{H_2}\approx1$ there and that the ratio $x_\mathrm{He}/x_\mathrm{H_2}$ remains constant throughout the atmosphere.} ($\xHetop$) we used are listed in table~\ref{tab:planets}.

\begin{figure*}[htbp] 
 \sidecaption\centering
\subfigure{ \includegraphics[scale=\sizefig,trim = 0cm 1.1cm 0.cm 0.cm, clip]{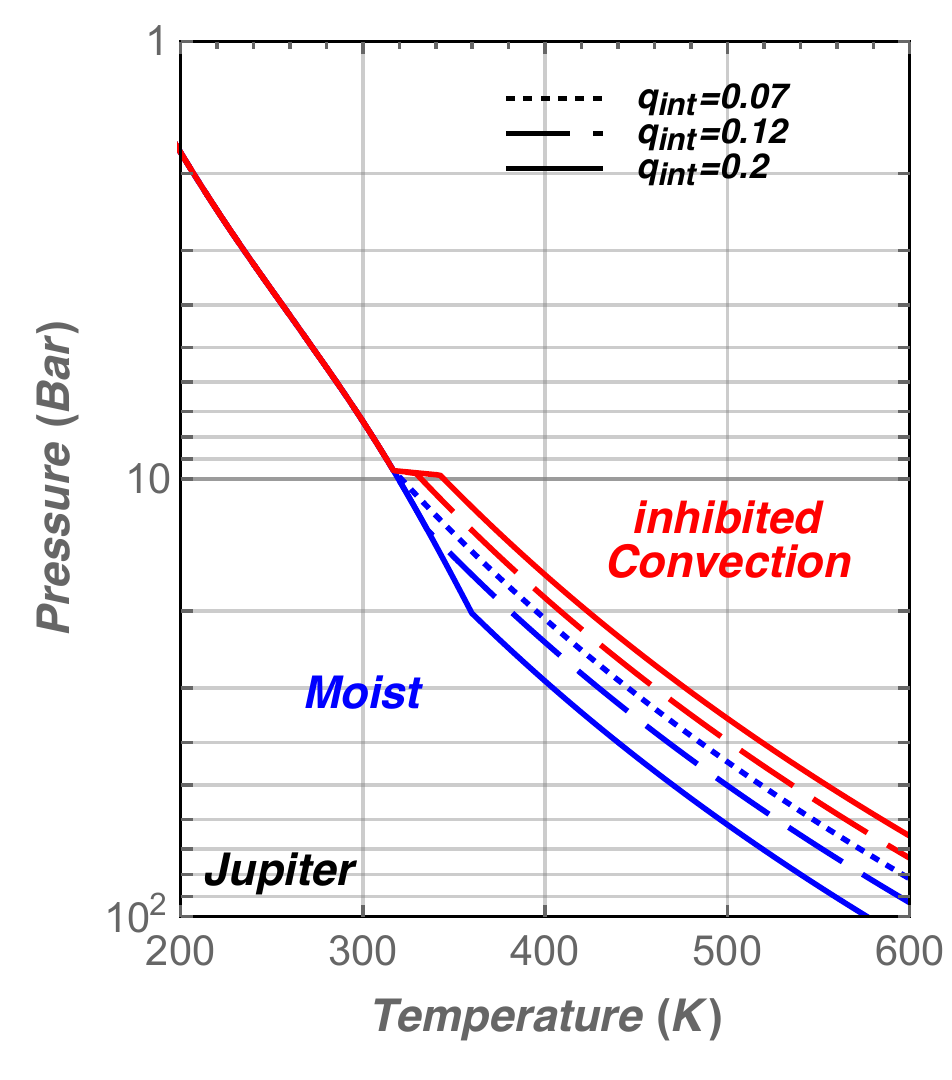} }
\subfigure{ \includegraphics[scale=\sizefig,trim = 1.1cm 1.1cm 0.cm 0.cm, clip]{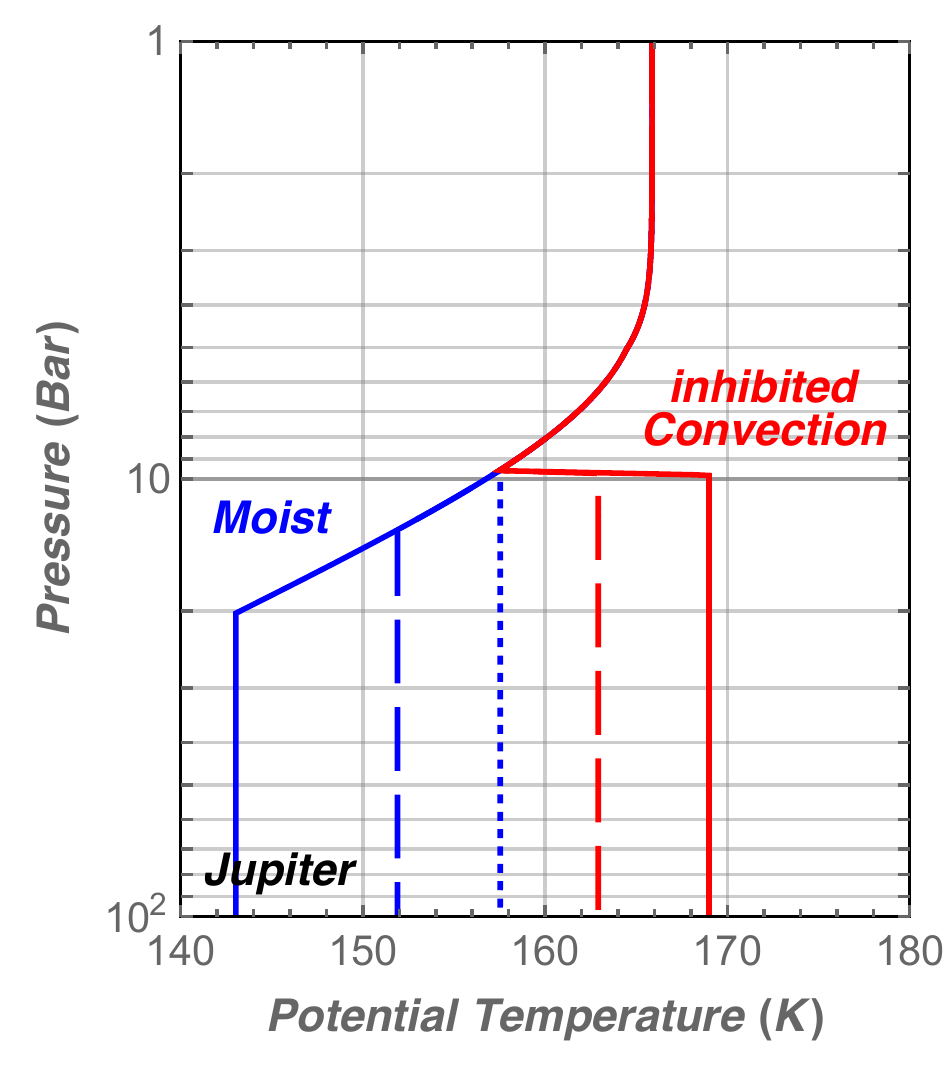} }
\vspace{-0.25cm}
\subfigure{ \includegraphics[scale=\sizefig,trim = 1.10cm 1.1cm 0.cm 0.cm, clip]{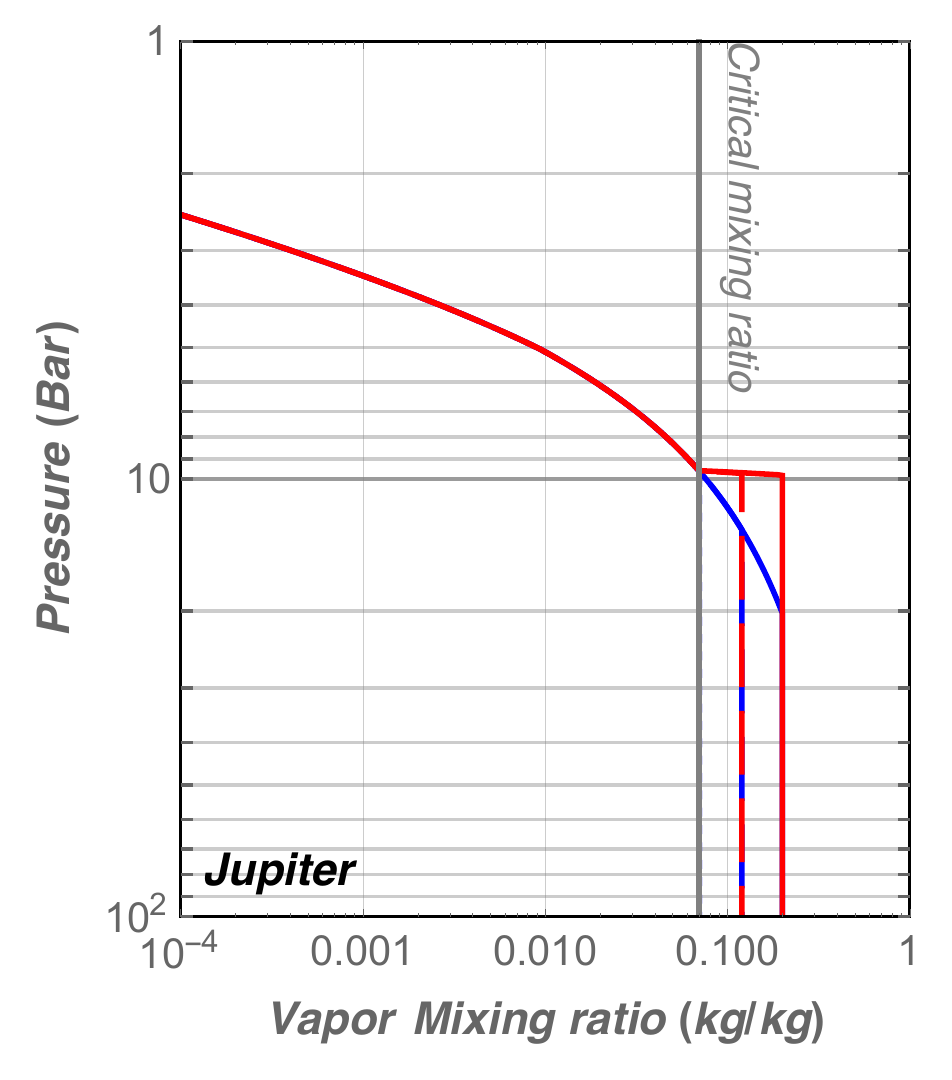} }
\subfigure{ \includegraphics[scale=\sizefig,trim = 0cm 1.1cm 0.cm .25cm, clip]{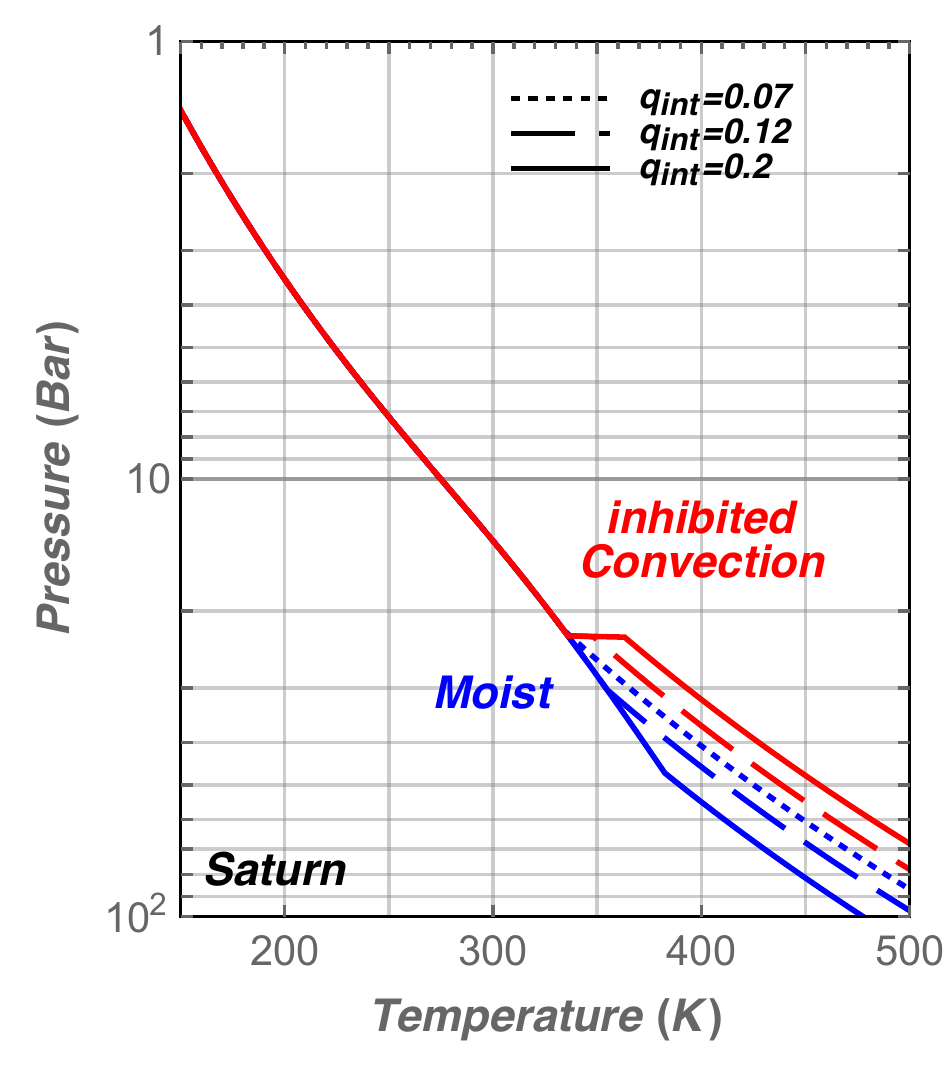} }
\subfigure{ \includegraphics[scale=\sizefig,trim = 1.1cm 1.1cm 0.cm .25cm, clip]{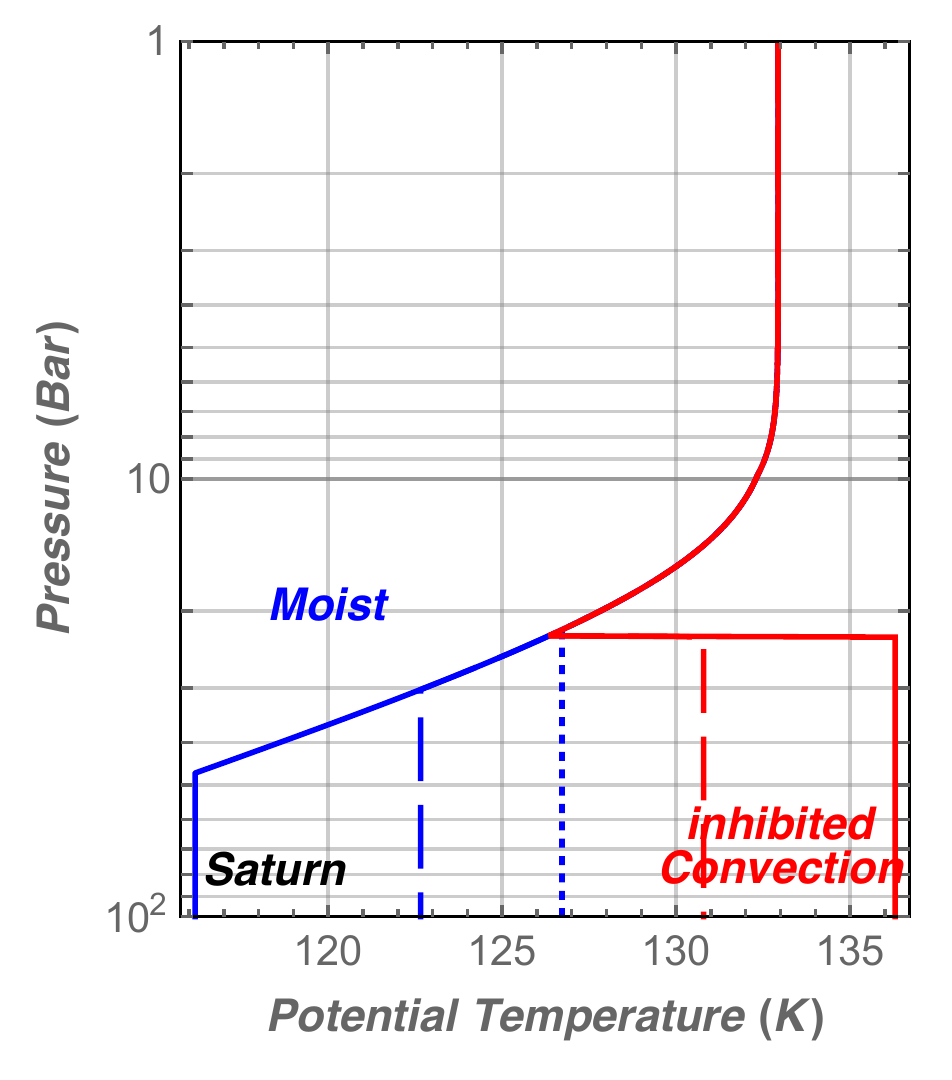} }
\vspace{-0.25cm}
\subfigure{ \includegraphics[scale=\sizefig,trim = 1.10cm 1.1cm 0.cm .25cm, clip]{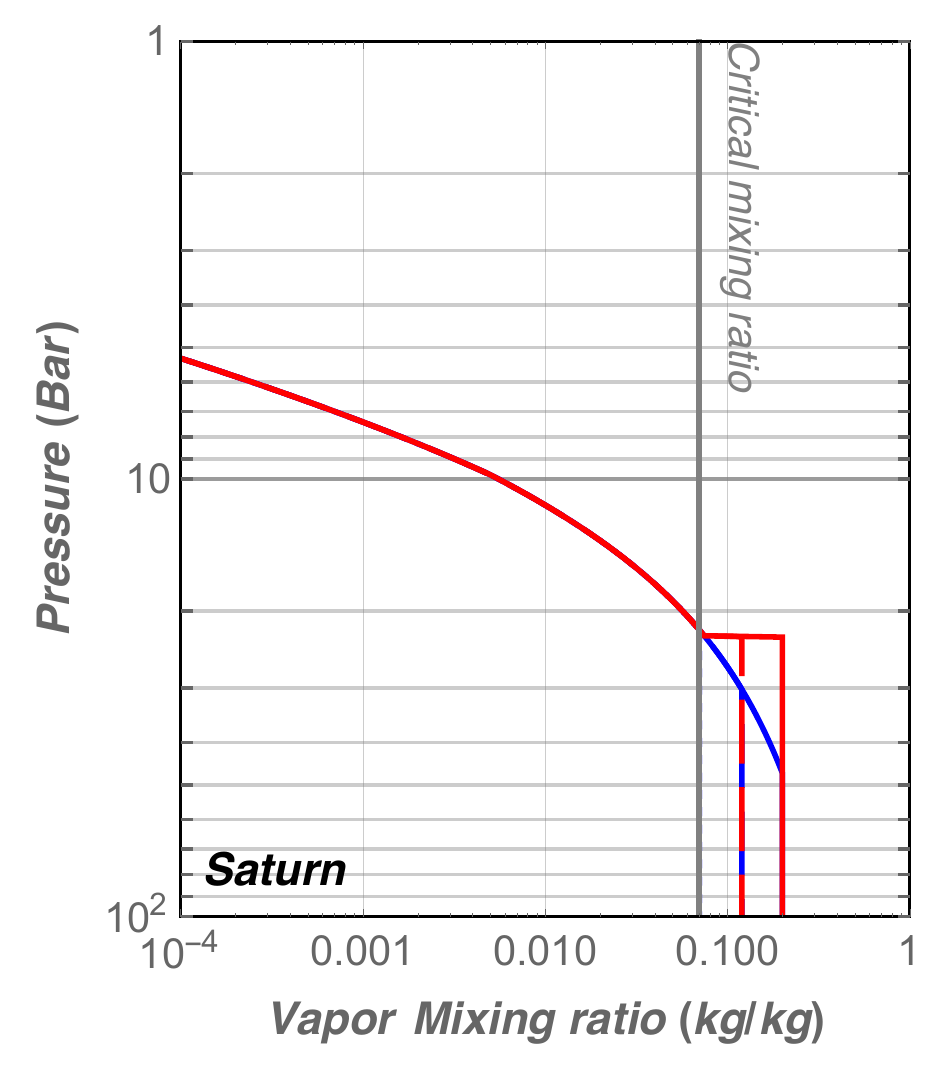} }
\subfigure{ \includegraphics[scale=\sizefig,trim = 0cm 1.1cm 0.cm .25cm, clip]{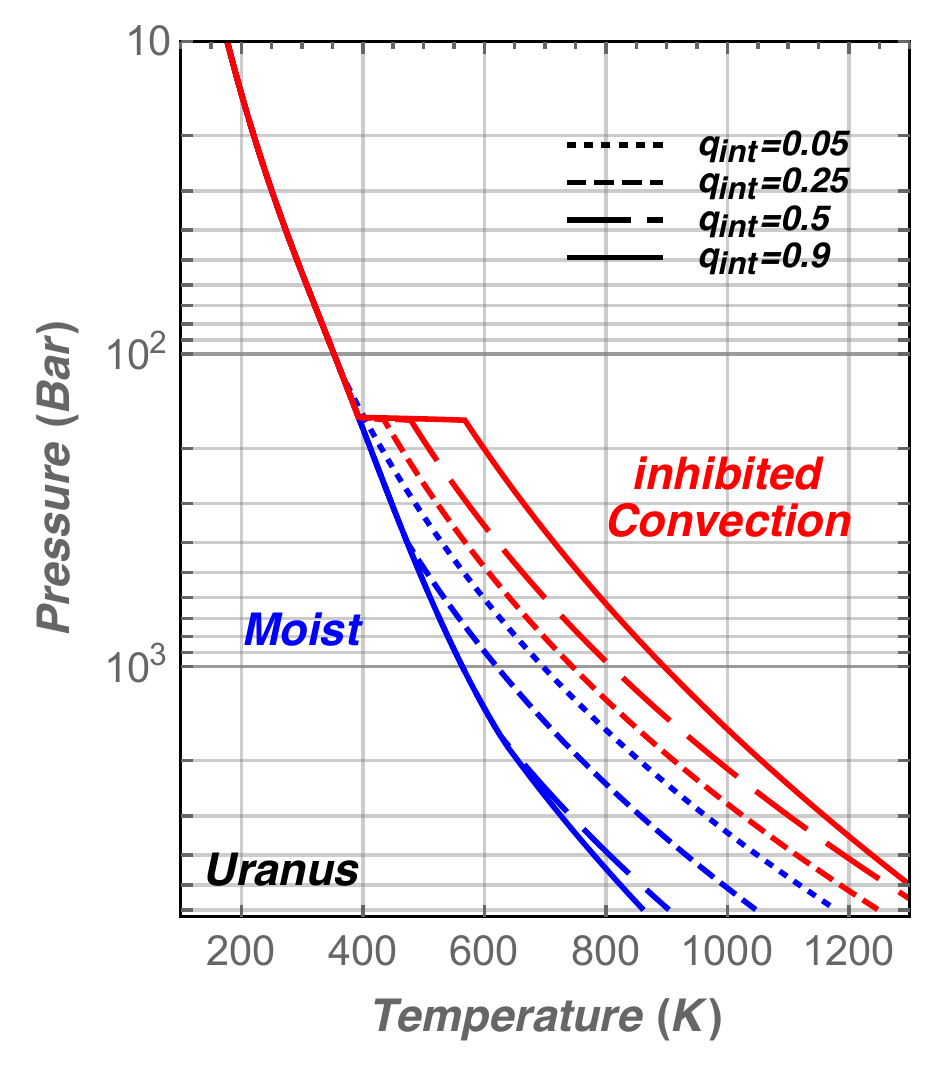} }
\subfigure{ \includegraphics[scale=\sizefig,trim = 1.1cm 1.1cm 0.cm .25cm, clip]{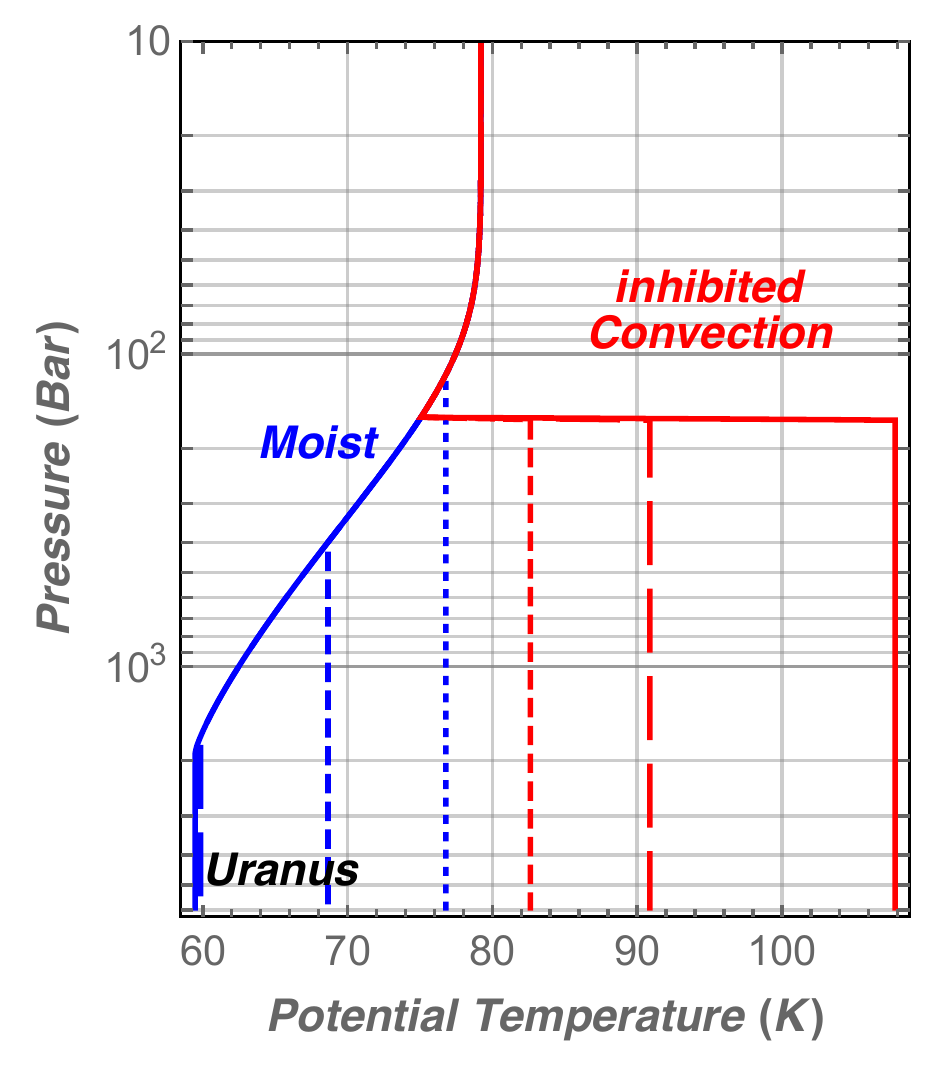} }
\vspace{-0.25cm}
\subfigure{ \includegraphics[scale=\sizefig,trim = 1.10cm 1.1cm 0.cm .25cm, clip]{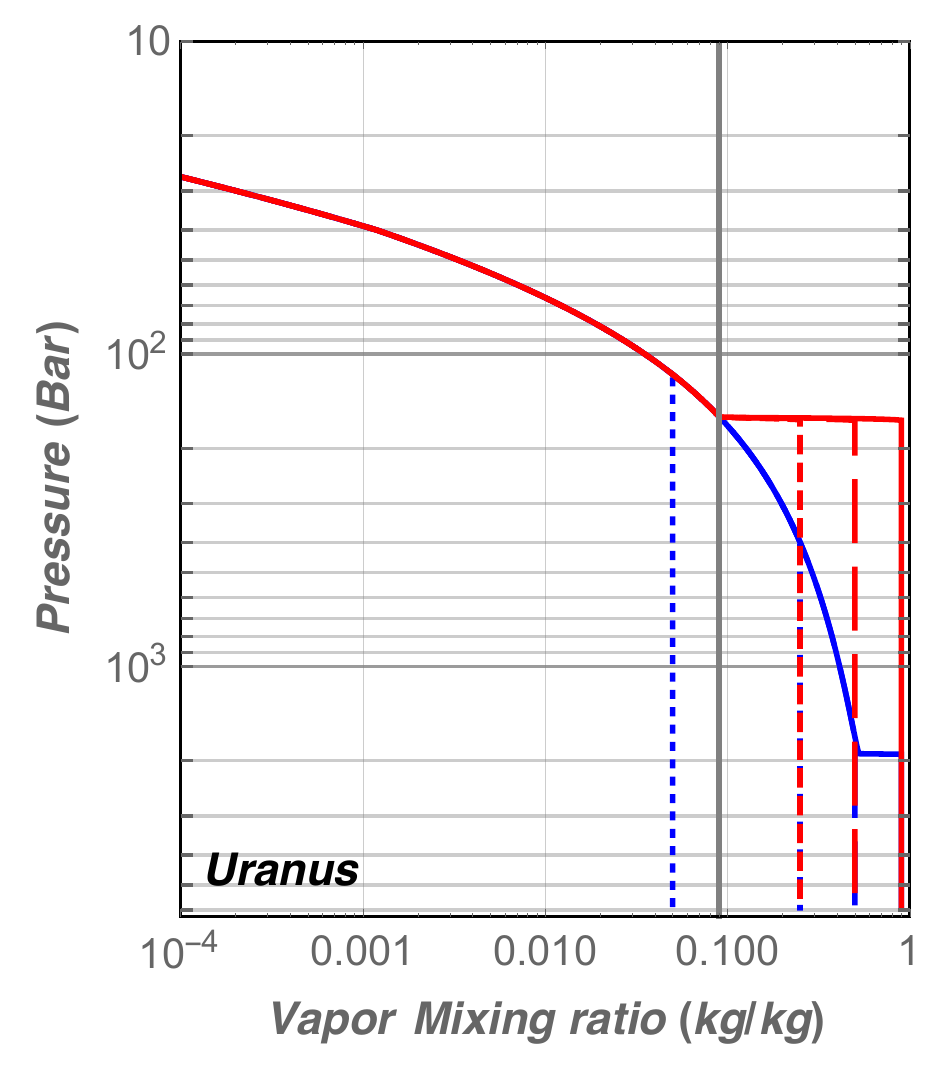} }
\subfigure{ \includegraphics[scale=\sizefig,trim = 0cm 0cm 0.cm .25cm, clip]{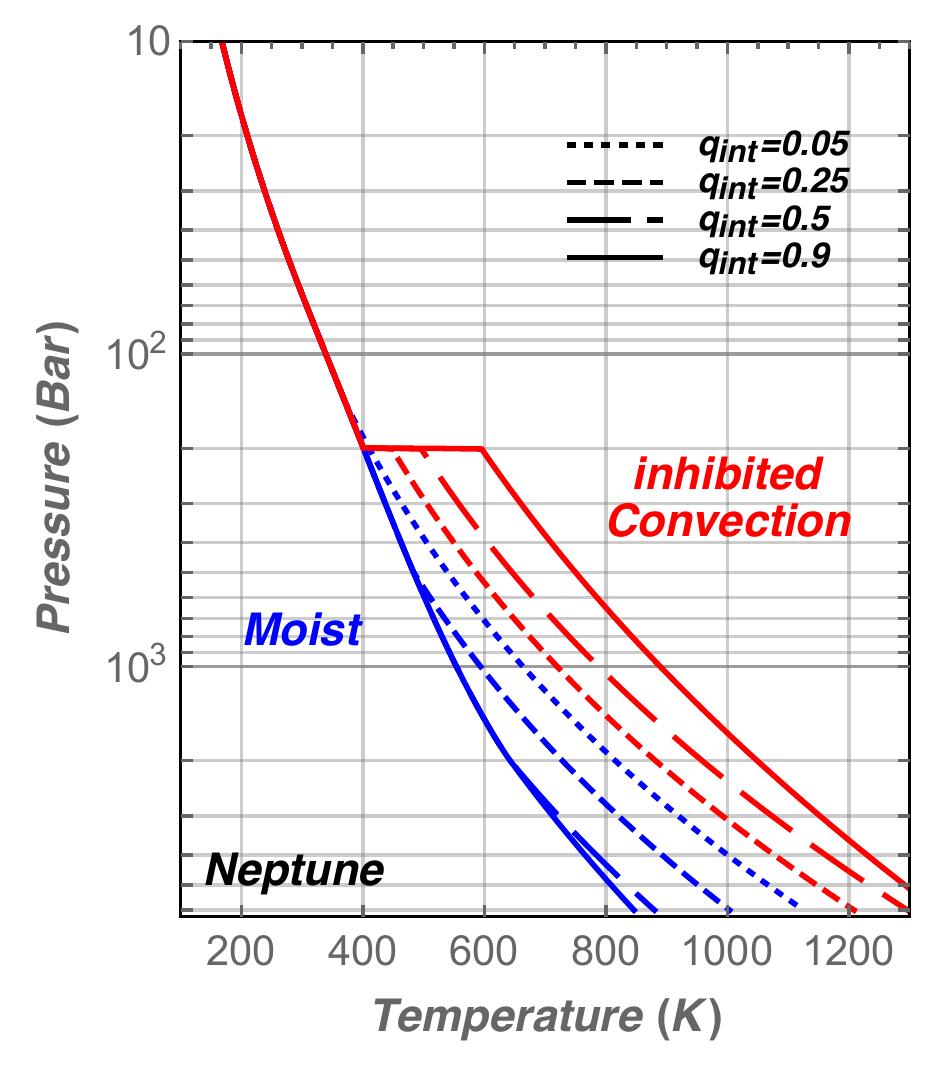} }
\subfigure{ \includegraphics[scale=\sizefig,trim = 1.1cm .cm 0.cm .25cm, clip]{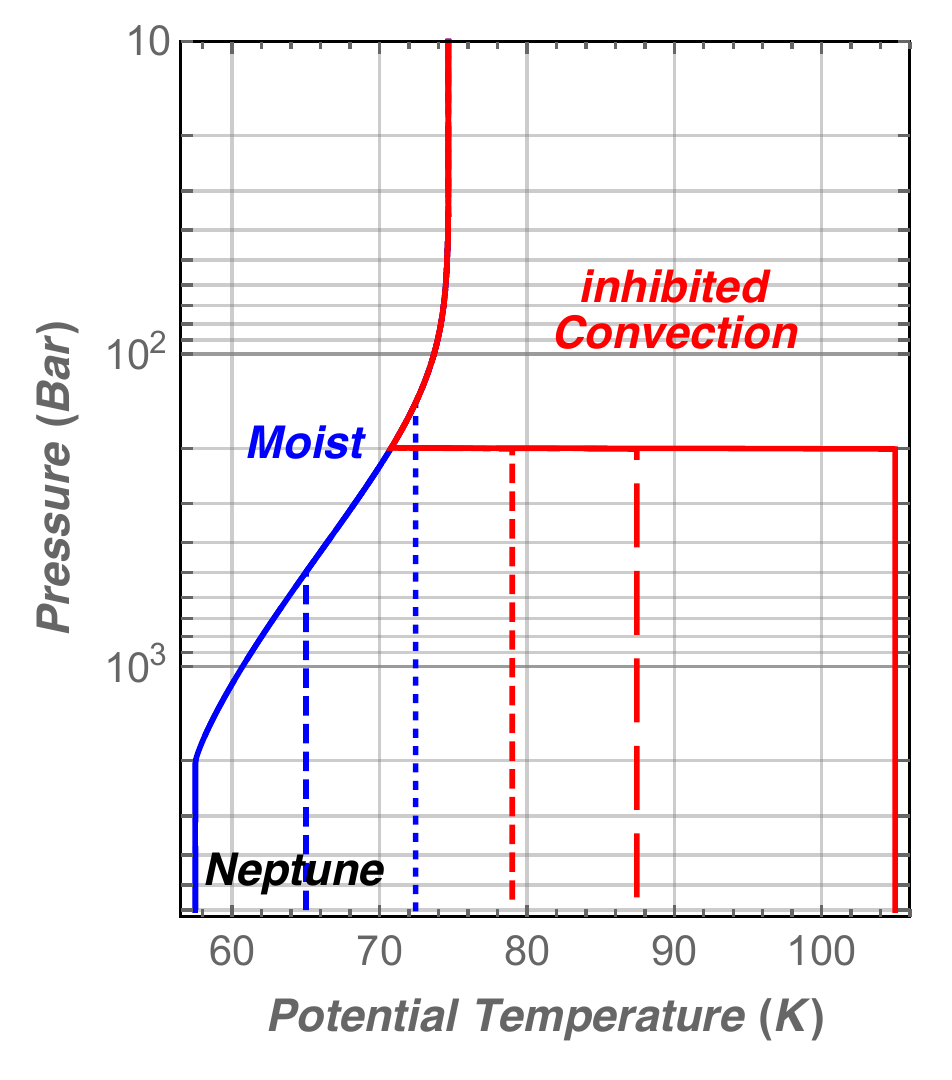} }
\subfigure{ \includegraphics[scale=\sizefig,trim = 1.10cm .cm 0.cm .25cm, clip]{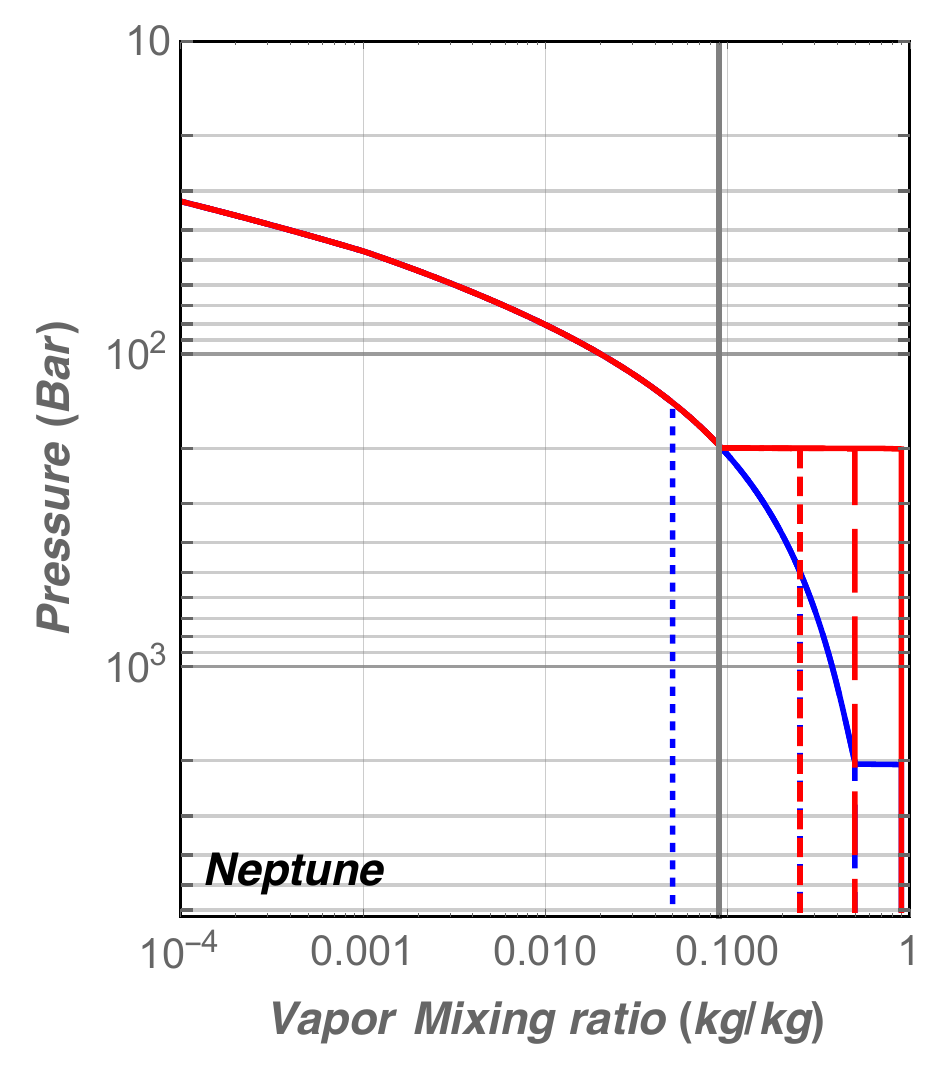} }
\vspace{-0.6cm}
\caption{
Temperature (left), potential temperature (middle), and vapor mixing ratio (right) profiles for the four giant planets (top to bottom; see label). For each planet, several internal water contents are shown (see legend on each line), corresponding to black stars in \fig{fig:deeptemperature}. Usual 2-layer models following a simple moist adiabat are drawn in blue, and 3-layer models with convection inhibition are shown in red (but only for models with $\qint>\qcri\sim$0.07 that actually possess an inhibition layer). The potential temperature accounts for variations in $\mu$ and $\cp$ in the adiabatic index (see \eq{pottemperature}), so that dry adiabats appear as vertical lines in the middle panel.}
 \label{fig:profile4planets}
\end{figure*}

\subsection{Results}

\begin{figure*}[htbp] 
 \centering
\subfigure{ \includegraphics[scale=.8,trim = 0cm 1.cm .8cm 0.5cm, clip]{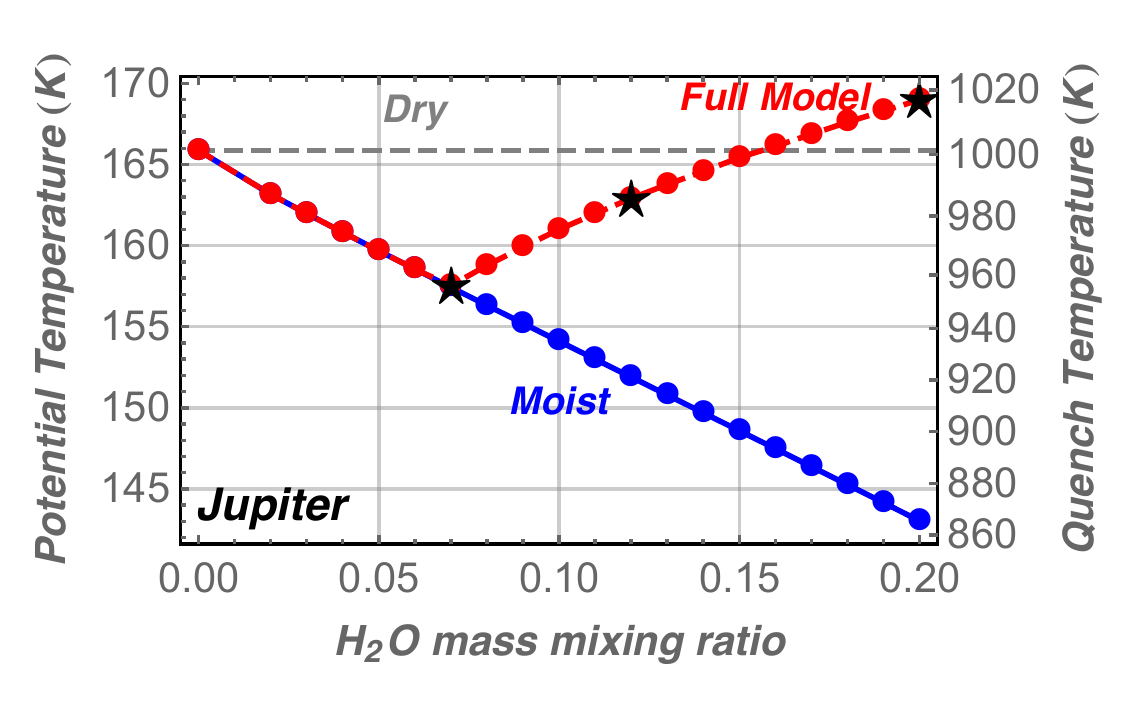} }
\subfigure{ \includegraphics[scale=.8,trim = .8cm 1.cm 0.cm 0.cm, clip]{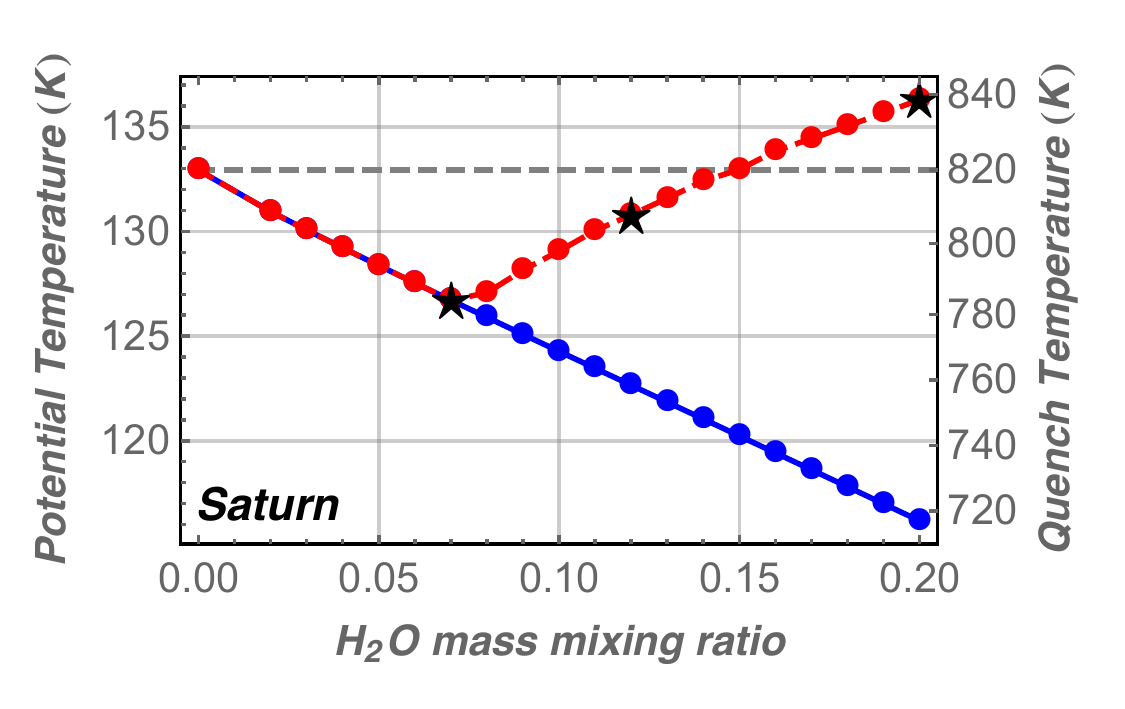} }
\subfigure{ \includegraphics[scale=.8,trim = 0cm .cm .8cm .5cm, clip]{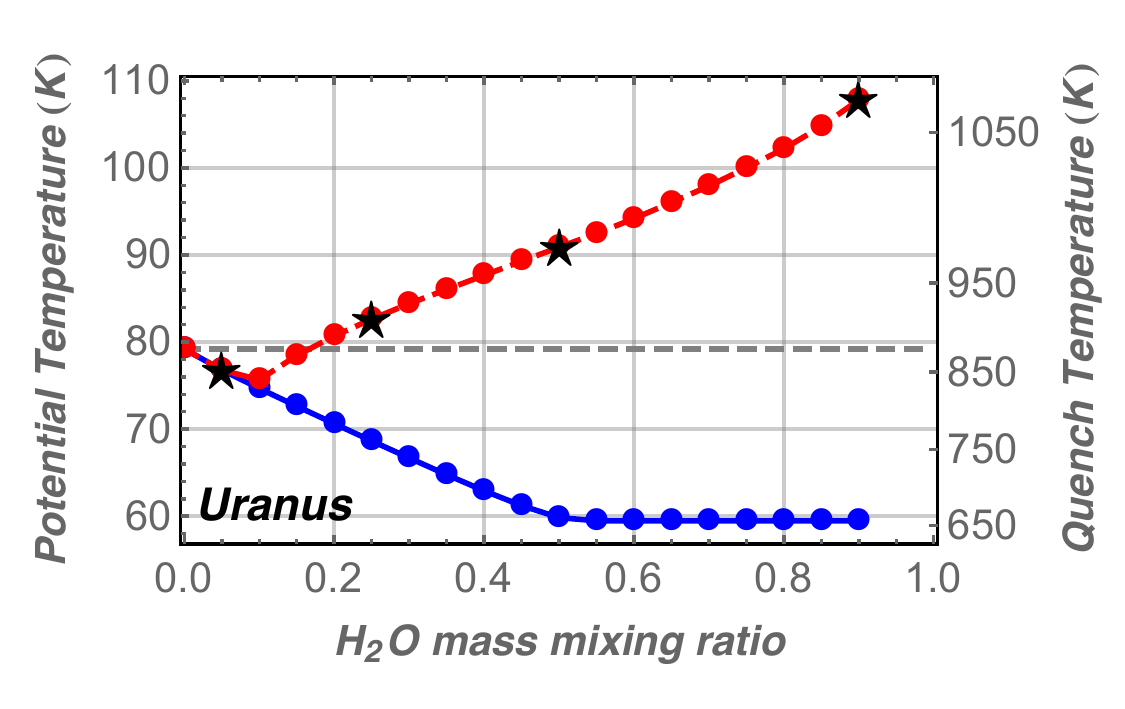} }
\subfigure{ \includegraphics[scale=.8,trim = .8cm .cm 0.cm 0.5cm, clip]{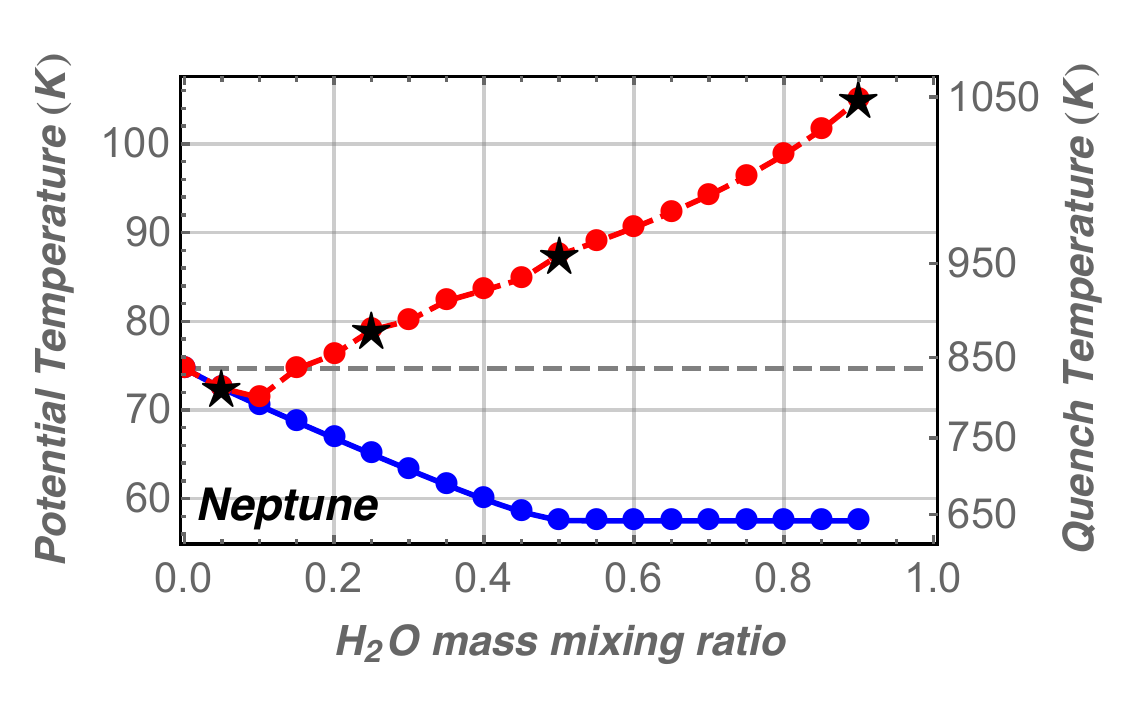} }
\caption{
Potential temperature of the deep adiabat of the four planets as a function of the internal water mixing ratio. Models taking into account the \mmw effect are shown with a red dashed curve. \textit{Black stars} highlight the actual models shown in \fig{fig:profile4planets}. The \textit{dashed gray} horizontal line gives the reference for a dry adiabat, and {blue dots} show the temperature expected for a usual 2-layer moist adiabat. The scale on the right hand side of each plot yields the conversion in real temperature at the pressure level roughly corresponding to the quenching of CO inside each planet ($\sim$400\,b for Jupiter and Saturn and $\sim$2000\,b for Uranus and Neptune)}
 \label{fig:deeptemperature}
\end{figure*}

The thermal profiles obtained for our four giant planets are shown in \fig{fig:profile4planets}. For small water enrichment, in the sense that $\qint<\qcri$, the atmosphere is separated in only 2 layers. In this regime, the potential temperature of the deep adiabat decreases when the internal water content increases (see \fig{fig:deeptemperature}). This is because  the potential temperature along a moist adiabat \textit{decreases} with depth, as can be seen in the middle panel of \fig{fig:profile4planets}.

As advertised, when the water content of the planet exceeds the critical threshold, a stable radiative layer develops. Because of the relatively large opacity at depth, the radiative gradient is one to two orders of magnitude larger than the adiabatic one, so that the radiative layer almost appears as a temperature jump. 
Counter-intuitively, even with a radiative layer, the potential temperature of the deep adiabat of a moist atmosphere is not necessarily higher than the potential temperature of a water-poor one. This is because the potential temperature along a moist adiabat \textit{decreases} with depth, as discussed above. However, at a given internal water mixing ratio, the deep adiabat is always hotter when a radiative layer is present compared to the usual parametrization of a moist atmosphere, as visible on \fig{fig:deeptemperature}.

The quantitative extent of this warming strongly depends on the internal mixing ratio of water. On one end, for Jupiter and Saturn, the heavy element abundance suggested by previous data are not expected to be more than 10-20\% in mass. Therefore, we decided to focus our analyses on the range $\qint\in \left[0,0.2\right]$. As this does not exceed the critical mixing ratio ($\approx0.07$) by a large factor, the warming remains modest, even if it can reach 20\,K in potential temperature, or about 150\,K at the approximate CO quenching level ($\sim 400\,$bar; \citealt{FMC09,VMS10}) as shown in \fig{fig:deeptemperature}. If there is no evidence for the existence of a radiative layer inside Jupiter, it can not be ruled out at the moment. For Saturn, however, if the recurrence of giant storms is indeed explained by convection inhibition near the cloud base, as argued by \citet{LI15}, such a radiative layer must exist. In fact, our critical water abundance, $\qcri$, is, in essence, the same criterion as the one used by these authors to determine the occurence of storms.

For Uranus and Neptune, on the other end, enrichments are believed to be much higher\footnote{This actually causes an interesting behavior with the highly enriched, usual 2-layer models. Indeed, in this case, water pressure in the atmosphere can reach the critical pressure before being in the dry region. This stops the moist adiabat as seen on \fig{fig:profile4planets}, and sets a lower limit on the potential temperature in the deep adiabat (see \fig{fig:deeptemperature}). Below that, the limit on the water vapor fraction of the atmosphere is set by miscibility constraints, but we will not treat that aspect. Anyway, this is not a problem in the more realistic 3-layer case because the dry region is reached much higher because of the higher temperature.}. For these planets, there is thus little doubt that some convection inhibition is at play near the water condensation region, just like what is observed near the methane cloud region \citep{Gui95}. In addition, the high enrichment can cause the warming to reach up to 50\,K in potential temperature. 
At the approximate depth of CO quenching ($\sim$2000\,bar; \citealt{CML14}), this amounts to more than a 400\,K warming compared to the usual moist adiabat (see \fig{fig:deeptemperature}).

Another prediction of our model is that, once $\qint>\qcri$, the depth of the cloud base is almost independent of the water~content---contrary to what is found for usual moist profiles (see \fig{fig:profile4planets}). 

\begin{table}
\centering
\begin{tabular}{llcccc}
\hline
&& Jupiter	 & Saturn & Uranus & Neptune \\
\hline
$\xHetop$ && 0.136 & 0.12 & 0.152 &  0.150 \\
$\temp_\mathrm{top}$&K& 166& 133&102& 97 \\
$\press_\mathrm{top}$&bar& 1.&1.&2.&2. \\
$\Fint$& W/m$^2$ & 5.4 & 2.6$^{a}$ & 0.04 & 0.41\\
\hline
\end{tabular}
\caption{Helium volume mixing ratio, internal heat flux ($\Fint$), and top temperature and pressure used for the giant planets and icy giants of the solar system. Except specific mentions, data are from \citet{Gui05}. For Uranus and Neptune, we decided to start our integration below the methane cloud base to avoid further assumptions. However, computed potential temperatures always use a 1\,bar reference pressure. $a$: \citet{LCG10}.}
\label{tab:planets}
\end{table}

%

\section{Discussion}

\subsection{Jupiter and the Galileo probe measurements}

In 1995, the Galileo probe made unique, \textit{in situ} measurements in Jupiter's atmosphere down to 22\,bars and a temperature of 428\,K. Even though the probe fell into a 5-micron hot spot, a region that appears quite different from other more \textit{average} locations on the planet, these measurements are to be considered to put our approach in perspective. 

In most of the tropospheric descent, that is between about 4 and 16 bars, the probe measured a temperature lapse rate of $-2$\,K/km that is consistent with a dry adiabat. Deeper, the gradient decreased in absolute value to reach about $-1.5$\,K/km near the end of the descent \citep{SKK98,MSY02}. 

The dry adiabat measured by the probe is not inconsistent with the possibility that the temperature profile may be either sub- or super-adiabatic elsewhere. However, the fact that the last measurements indicated a sub-adiabatic gradient for pressures larger than 16\,bar can be interpreted in two ways. One possibility is that the opacities are low enough in this region to allow for a direct radiative cooling as obtained by \citet{GGC94} when not including the opacity of water. This interpretation, discussed by \citet{SKK98} and \citet{MSY02}, would have to be tested with modern opacity data and using the water abundance measured by the Galileo probe. If on the other hand the opacity is large enough to ensure the atmosphere to be convectively unstable according to the Schwarzschild criterion, this sub-adiabatic gradient must be due to the fact that the atmosphere surrounding the hot spot is at lower temperatures. Given that the horizontal extent of a hot spot is much larger than its vertical extent we would then conclude that the probe reached the bottom of the hot spot and that both the temperature and water abundance started to converge to their mean atmospheric value \citep[see e.g.,][]{SI98}. 

Quantitatively, the offset in temperature between the probe measurement at 22\,bar and the dry adiabatic prediction was about 4\,K \citep{SKK98}. In the framework of the second interpretation, we would conclude that the environment of the hot spot is at least 4\,K cooler than the hot spot itself, but that this offset was caused by water condensation ensuring a moist adiabatic profile at significantly lower pressures. At 22 bar, this corresponds to an offset in potential temperature of 1.5\,K (assuming a zero offset between the hot spot and the environment at 1\,bar). As shown by \fig{fig:jupiter_zoom} below, this corresponds to a minimum amount of water of $\qint\approx$ 0.01-0.02, i.e., 1.5-3 times the solar value. A maximum amount can also be derived from \fig{fig:deeptemperature}, at $\qint\approx$ 0.13, i.e., about 18 times solar. Because the water abundance measured by Galileo was still significantly below these values, it is likely that the bottom of the hot spot had not been reached and that the temperature offset is significantly larger than 4\,K, resulting in higher water abundances. 

If the second interpretation is correct, we can put important constraints on the abundance of water in Jupiter's deep atmosphere. Whether this is the case will be directly tested by \textit{Juno}'s radiometric measurements \citep[see ][]{PJO08,DSD14}.

\begin{figure}[htbp] 
 \centering
\subfigure{ \includegraphics[scale=.6,trim = 0cm .cm .cm 0.cm, clip]{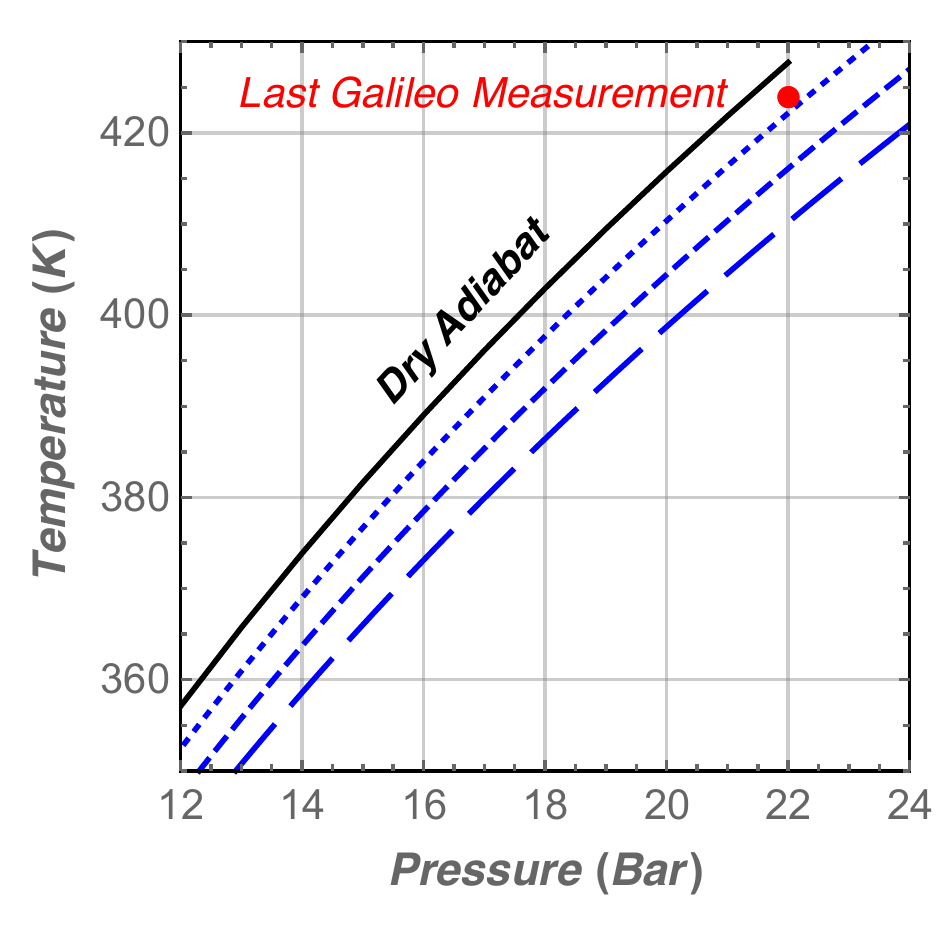} }
\caption{
Temperature profile for Jupiter focused on the end of the Galileo descent (near 22\,bar). The black solid line is the dry adiabat and the blue curves are moist adiabat with $\qint=$ 0.02 (dotted), 0.04 (dashed), and 0.06 (long dashed). The red dot represents the Galileo measurement at 22\,bar, which is 4\,K colder than the dry adiabat. Profiles here may be slightly warmer than in \figs{fig:profile4planets}{fig:deeptemperature} because they were adjusted to have a 260\,K temperature at 4.18\,bar to be consistent with the analysis of \citet{SKK98}.
}
 \label{fig:jupiter_zoom}
\end{figure}

%

\subsection{Time variability and relation to Saturn giant storms}

By construction, our method to build model atmospheres is time independent and assumes thermal equilibrium. This is the reason why, if convection inhibition occurs, a layer with a very steep thermal gradient must be present to carry out the internal flux of the planet radiatively. 

\citet{LI15} envisioned a \textit{seemingly} very different scenario. In their model, convection inhibition completely stops energy exchange between the interior and the upper atmosphere which leads to the cooling of the latter. The internal flux is thus carried out only episodically during short outbursts---giant storms. In this scenario, no \textit{permanent} radiative layer is needed, and these authors thus disregard the possibility of the internal adiabat being much hotter than commonly thought. 

There are, however, a few limitations to their picture. The major one is that, if the internal water mixing ratio is sufficient to inhibit convection at a given latitude, convection should be inhibited at all latitudes\footnote{Except maybe if the water mixing ratio at depth is incredibly close to the critical one, or show large, unpredicted horizontal inhomogeneities.}. Thus, energy would be brought upward only through giant storms\footnote{Smaller storms are which happen much more often are expected to originate in the moist troposphere and not to reach the deep adiabat}. This means that each storm would have to carry the internal flux accumulated on an entire cycle and for the \textit{whole} planet---or a large fraction of it. The total accumulated internal flux over the planet is about 10$^{26}$\,J for a 30 year cycle. A rough estimate of the energy released by the 2010 great white storm, $E_\mathrm{GWS}$, can be obtained considering that the storm caused an average $\Delta T\approx 3$\,K warming over the whole longitude circle between 30$^\circ$ and 40$^\circ$ latitude ($A\approx 3\times10^{15}$\,m$^2$) down to a pressure of 100-400\,mbar \citep{FHI11,AGC14}. This yields 
\balign{E_\mathrm{GWS}=A\frac{\press}{g}\cp \Delta T\approx 5\times10^{23}\mathrm{J}.}
Although this does not consider latent heat stored by water vapor, this still is a couple orders of magnitude too weak. 

This seems to suggest that a more continuous process is at play to release internal energy. A radiative layer as the one described in previous sections could play this role. To investigate this possibility, we made an attempt to develop a full 1D time-stepping model of giant planets atmosphere incorporating a realistic water cycle and radiative transfer based on the LMD climate model for Saturn \citep{GSS14}. This proved highly difficult because of the high aspect ratio between the shortest and the longest timescale to resolve: the convection timescale which sets time-steps to be smaller than an hour and the thermal evolution of the deep atmosphere which takes hundreds---if not thousands---of years to reach equilibrium. This was further complicated by the vertical resolution needed to model the thin radiative layer. In practice, when the resolution was locally high enough to resolve the temperature jump, the model was numerically unstable for practical time-steps. \textit{In fine,} the model did not run under fully realistic conditions, hence our decision not to report all the details of this experiment here. However, some tests under more idealized conditions revealed interesting trends that might shed some light on the matter at hand. Therefore, please bear in mind that the following findings should be regarded as tentative and merely suggestive. 

In these simple simulations, the main difference with the usual 1D LMD climate model \citep{WFS10,GSS14} is that both dry and moist convective adjustments criteria were changed according to \sect{sec:conv_adiabatic} to take into account \mmw effects. Similarly to the results of \citet{LI15}, we found that under some conditions on the subsaturation of the moist troposphere, a storm cycle could develop. Yet, during the long cooling phases, energy coming from the interior would accumulate below the inhibition layer, slowly forming a temperature jump. This deep temperature increase would end when the radiative flux through the stable layer would equal the internal flux, as expected. An interesting trend is that storms did not appear when saturation was more strictly enforced---for example by forcing a very efficient re-evaporation of falling precipitations. In this case, a radiative layer formed, and the atmosphere above it settled in a quasi steady-state where small-scale moist convection would carry the flux. 

This leads us to think that there is no contradiction between the two pictures drawn hereinabove. In fact, combining the two might lead to a more consistent picture. Because of the \mmw jump at the water cloud base, Saturn's atmosphere could exhibit a temperature jump at this depth. Over most of the atmosphere, this could release internal heat and sustain the global infrared excess observed. At the same time, in some specific regions---especially dynamically subsaturated regions---, a small energy imbalance could still develop, powering episodic giant storms originating just above the stable radiative layer. 

This also solves another limitation of the scenario presented by \citet{LI15}. The timescale they find for the cooling phase is in large part determined by the depth of the cloud base---that they take to be $\sim$\,20\,bar (see bottom of page 399). But the depth of the actual cloud base depends on the actual water mixing ratio in a standard moist adiabatic atmosphere. In our profiles derived from more realistic thermodynamical data, the cloud base reaches 20\,bar around $\qint\sim\qcri$. Notwithstanding the admitted roughness of their estimate, the coincidence is troubling and it would seem odd---although not impossible---that Saturn's water content would be close to that value by chance. 

In our scenario, the conditions for the existence and location of a stable layer could explain this coincidence. Indeed, the top of the radiative layer is determined by the fact that $\qvap=\qcri$, and this stable layer actually decouples convection above it from convection below. When $\qint>\qcri$, the location of the stable layer thus becomes independent of $\qint$ as shown in \fig{fig:profile4planets}. Unfortunately, this prevents us to draw any inference on Saturn's internal water ratio based on the recurrence period of the giant storms, except that it exceeds the critical threshold. On the bright side, the presence and thickness of the radiative layer could have major impacts on the winds and wave propagation properties in the deep atmosphere. These may be constrained by high-order gravitational potential measurements.

\subsection{Storms on other planets?}

As discussed by \citet{LI15}, the absence of giant storms on Jupiter could be taken as an argument for a sub-critical water mixing ratio at depth---although this does not mean that the whole jovian gaseous envelope should be water poor \citep{LC12}. 

While quite convincing, this argument does not have to be definitive. The occurence of Cronian storms only at specific locations and seasons \citep{SDD12} indeed suggests that specific dynamical conditions and seasonal forcing might be necessary. Compared to Jupiter, Saturn, which presents a slightly larger eccentricity and a much higher obliquity, and which possesses shading rings, exhibits a much stronger seasonal cycle. Add to that the fact that the cloud base is located at lower pressures and is thus thicker (because the radiative gradient is lower), which might have an effect as well.

With their strong enrichment---and to the extent that the occurence of giant storms would be only conditioned by the deep atmospheric water content---Uranus and Neptune would be obvious targets to look for giant water storms. We might therefore ask whether or not the fact that we did not detect any such event poses any constraint. The thing is that considering the rough storm recurrence period estimate made by \citet{LI15}---and factoring in the lower flux and much higher pressure for the cloud deck ($\sim$200\,bar)---we should expect storms to occur about 100 times less frequently than on Saturn. This makes it very unlikely that one happened since humankind developed the instruments needed to spatially resolve the outer planets. 

An intriguing possibility, though, would be that these storms would actually happen and be global in scale---because of the lower Rossby number. In this instance, the infrared excess released by the planet could vary periodically. The dichotomy between the measured internal fluxes for Uranus and Neptune might then simply be explained by the fact that we might be looking at two planets in very different phases of this cycle. Although not quite possible at the moment, 3D simulations of the deep atmospheric dynamics of these planets on long timescales would be needed to investigate this possibility. 

In the meantime, convection inhibition due to methane has also been reported on Uranus and Neptune \citep{Gui95}. Because methane clouds are located at much smaller depth (1-2\,bar) the giant storms they cause should also happen with a period of a few decades. Storms have indeed been reported on Uranus \citep{DSF14}. But whether are not these storms are related to Saturn's great white storm, or if they occur with a period independent of the orbital period remains to be clarified.

\subsection{Implications for abundance retrieval}

If a radiative layer is present in the deep atmosphere of the giant planets, this strongly puts in question the results from past and future chemical retrieval procedures. 

For indirect methods, first, when retrieved molecular abundances in the observable atmosphere are combined with thermochemical models to infer the elemental abundance at depth, an assumption on the unobserved, deep temperature profile has to be made. Without further constraints, the usual 2-layer moist then dry adiabat prescription is often used. As shown above, this may strongly underestimate the temperature at the depth at which the quenching of key species, like CO, occurs (up to 150\,K in Jupiter and Saturn, and 400\,K is Uranus and Neptune as visible in \fig{fig:deeptemperature}). As a result, the relation between the expected molecular abundance and the elemental abundance is changed. 

Fortunately, as the only free parameter of our model is the deep water mixing ratio, there is a relation between the temperature jump and the oxygen abundance in the interior. Although time and spatial variability might be an issue, in principle, it is thus possible to build completely self-consistent atmosphere models accounting for convection inhibition without adding any free parameter. New relations between the abundances measured in the atmosphere and those inferred at depth should be calculated. 

The temperature increase at depth caused by the presence of a stable layer should also impact retrieval methods themselves. Indeed, instruments aiming at directly measuring the water abundance inside giant planets may also have to make some assumptions on the temperature profile in the region probed \citep{BS89,PJO08,DSD14,CPG15}. There again, care should be taken to use a completely self-consistent profile if the deep water abundance retrieved is greater than the critical ratio.

\section{Conclusion}

Because it is not observable with current techniques, assumptions on the thermal profile in the deep atmosphere of giant planets are key in many applications. Here, we have shown that the commonly used prescriptions---a moist adiabat underlain by a dry one---is probably incorrect for at least three of our four Solar System giant planets.

Indeed, by performing a linear analysis, we have confirmed that the convection inhibition mechanism described by \citet{Gui95} is still efficient when diffusion of both heat and vapor are allowed. We further demonstrated that when condensation is efficient, it suppresses the double-diffusive instability, leaving only radiative processes to transport energy through a layer undergoing convection inhibition. This thus led us to posit the existence of a stable radiative layer near the cloud base of any condensing species whose deep abundance exceeds a critical threshold given by \eq{qcri}.

Although this has been observed with methane in the atmosphere of Uranus and Neptune \citep{Gui95}, the occurence of this process around the condensation level of water, the molecule with the greatest potential effect, has never been directly observed. We thus developed atmospheric models of the four major planets to quantify the impact of water condensation on the thermal profile. This showed that usual prescriptions significantly underestimate the temperature at depth, sometimes by several hundreds of degrees for very enriched interiors. If Jupiter is sufficiently enriched, it will be very interesting to see whether or not the future missions to orbit the planet will be able to collect more direct evidence of a non-convective region where water clouds are expected. 

Although this has many different implications in our Solar System, it does not end there. The process of convection inhibition due to condensation should be ubiquitous in atmospheres with a relatively low background \mmw and potentially high abundances in condensing species: Saturn- and Neptune-like planets, terrestrial planets with some leftover primordial atmosphere, enriched brown dwarfs, etc. But condensation need not be the only process creating the \mmw gradient. Chemistry \citep{TAC16} or H/He demixing in the interior could do this as well.

\begin{acknowledgements}
JL thanks O. Venot, T. Cavalie, S. Guerlet, A. Spiga, and T. Fouchet for helpful discussions. JL is also indebted to S. Guerlet for making the radiative data for Saturn available. This project has received funding from the European Research Council (ERC) under the European Union's Horizon 2020 research and innovation programme (grant agreement n$^\circ$ 679030/WHIPLASH) 
\end{acknowledgements}

\appendix

\section{Some relations}\label{app:relations}

Here are a couple relations that are used in the main text. First, the definition of the mixing ratio entails 
\balign{\qvap=\frac{\rhovap}{\rhoair+\rhovap}=\frac{\Mvap\pvap}{\Mair\pair+\Mvap\pvap}=\frac{\epsilon \,\pvap}{\press+\left(\epsilon-1\right)\,\pvap},}
where $\epsilon\equiv \Mvap/\Mair$. The derivative is given by
\balign{
\dqvap&=\epsilon\, \frac{\left(\press+\left(\epsilon-1\right)\,\pvap\right) \,\d \pvap- \pvap \d \left(\press+\left(\epsilon-1\right)\,\pvap\right)}{\left(\press+\left(\epsilon-1\right)\,\pvap\right)^2} \nonumber\\
&=\epsilon\, \frac{ \pvap \press }{\left(\press+\left(\epsilon-1\right)\,\pvap\right)^2}\, \left(\d \ln \pvap-\dlp\right)\nonumber\\
&=\qvap^2\, \frac{\press }{\epsilon\pvap}\, \left(\d \ln \pvap-\dlp\right) \label{dqv}.}
Using definitions and \eq{defamu}, one can recover
\balign{\epsilon\frac{\pvap}{\press}&=\frac{\mum}{\Mair}\qvap=\frac{\qvap}{1-\mratio \qvap},}
and \eq{defgammasat}. To obtain values at saturations, one just needs to replace the vapor pressure and mixing ratio by their saturations values ($\left\{\pvap,\qvap\right\}\rightarrow\left\{\psat,\qsat\right\}$).

\section{Further considerations on moist convection}
\label{app:moistconvection}
\subsection{Effect of condensates mass loading}

With our convention, the total density is the sum of the gas density and the density of condensates (that are assumed to have a negligible volume), i.e. 
\balign{\rho=\rho_\mathrm{g}+\rho_\mathrm{c}\equiv\rho_\mathrm{g}\left(1+\rho_\mathrm{c}/\rho_\mathrm{g}\right). 
}
Again, with our convention (see \sect{sec:dry_processes}), $\rho_\mathrm{c}/\rho_\mathrm{g}\equiv \qcon,$ so that an infinitesimal variation of density compared to the mean state is given by
\balign{\label{EOSwithCondensates}
\d \ln \rho= \d \ln\press +\amu \,\d \ln\qvap -\d \ln \temp+ \d \ln \left(1+\qcon\right).}

In a clear, nearly saturated environment, i.e. just before convection and condensation occur, the density difference between two levels separated by $\d z$ is 
\balign{
&\left.\d \ln \rho\,\right|_\mathrm{env} = \nonumber \\
&\ \ \ 
=-\left(1 +\amu \,\left.\frac{\partial \ln\qsat}{\partial \ln \press}\right|_{\temp}+\amu \,\left.\frac{\partial \ln\qsat}{\partial \ln \temp}\right|_{\press}\frac{\d \ln \temp}{\d \ln\press} -\frac{\d \ln \temp}{\d \ln\press}\right)\,\frac{\d z}{\hp} \nonumber\\
&\ \ \ =-\left(1 -\mratio \qsat+\left(\amu \gammasat -1\right) \delt\right)\,\d z/\hp,
}
where \eq{defamu} has been used.
In the rising eddy, the difference is that i) the thermal gradient follows the moist adiabat, and ii) we now account for the retention of condensates forming on ascent, yielding
\balign{
\left.\d \ln \rho\,\right|_\mathrm{edd} 
&=-\left(1 -\mratio \qsat+\left(\amu \gammasat -1\right) \delmoist\right)\,\d z/\hp+\d \ln \left(1+\qcon\right).
}
So the unstability criterion derived from the density difference becomes
\balign{
0<\left.\frac{\d \ln \rho}{\d z}\,\right|_\mathrm{env}-\left.\frac{\d \ln \rho}{\d z}\,\right|_\mathrm{edd} }
i.e.
\balign{0<\left(1-\amu \gammasat \right)\left(\delt- \delmoist\right)-\hp\frac{\d \,\,\,}{\d z}\ln \left(1+\qcon\right).
}
Although we could go further and try to link the amount of condensates to the amount of vapor lost, it is clear from the negative sign of the last term in this expression that \textit{any amount of condensates retained during ascent tends to hamper convection} \citep{Gui95}.

\subsection{effect of subsaturation during subsidence}

We now turn our attention to a subsiding, i.e. sinking, eddy. Contrary to the main text, here we will not assume the presence of condensate able to sublimate and keep the vapor at saturation. During descent, the adiabatic cooling warms the gas. So, in essence, subsiding regions do not undergo condensation and the vapor mixing ratio stays fixed within a parcel.

Following the analysis above, the density change in the sinking eddy is thus 
\balign{
\left.\d \ln \rho\,\right|_\mathrm{edd} &=\left(\delad-1\right)\,\d z/\hp.
}
The condition for a sinking eddy to keep sinking in an environment that is near saturation on average is thus given by
\balign{0<\delt- \delad-\left(\amu \gammasat \delt-\mratio \qsat\right)\equiv\delt- \delad-\delmu,
}
which reduces, as expected, to the Ledoux criterion for a dry motion.

\bibliography{biblio}
\bibliographystyle{aa}



\end{document}